\documentclass[preprint,review,12pt]{elsarticle}
\usepackage{psfig,epsf,rotating,aalongtable,txfonts}
\usepackage{color}

\begin{document}
\begin{frontmatter}
\baselineskip=25pt
\textwidth 15.0truecm
\textheight 21.0truecm
\topmargin 0.2in
\headsep 1.2cm

\title{Spectroscopic Survey of X-type Asteroids\footnote{Based on observations carried out at the European Southern
Observatory (ESO), La Silla, Chile, ESO proposals 074.C-0049 and 078.C-0115, at the Telescopio Nazionale Galileo, La Palma, Spain, and at the Mauna Kea NASA IRTF telescope, Hawaii, USA.}}
\author{S. Fornasier$^{1,2}$, B. E. Clark$^{3}$, E. Dotto$^{4}$}

\maketitle
\noindent
$^1$ LESIA, Observatoire de Paris, 5 Place Jules Janssen, F-92195 Meudon Principal Cedex, France\\
$^2$ Univ. Paris Diderot, Sorbonne Paris Cit\'{e}, 4 rue Elsa Morante, 75205 Paris 
Cedex 13 \\
$^3$ Department of Physics, Ithaca College, Ithaca, NY 14850, USA \\
$^4$ INAF, Osservatorio Astronomico di Roma, via Frascati 33, I-00040 Monteporzio Catone (Roma), Italy \\
\noindent
Submitted to Icarus: December 2010\\
e-mail: sonia.fornasier@obspm.fr; fax: +33145077144; phone: +33145077746\\
Manuscript pages: 38; Figures: 9; Tables: 7 \\
\vspace{3cm}

{\bf Running head}: Investigation of X-type asteroids

\noindent

{\it Send correspondence to:}\\
Sonia Fornasier  \\
LESIA-Observatoire de Paris  \\
Batiment 17 \\
5, Place Jules Janssen \\
92195 Meudon Cedex \\
France\\
e-mail: sonia.fornasier@obspm.fr\\
fax: +33145077144\\
phone: +33145077746\\

\newpage
\vspace{2.5cm}

\begin{abstract}

We present reflected light spectral observations from 0.4 to 2.5 $\mu$m 
of 24 asteroids chosen from the population of asteroids initially classified 
as Tholen 
X-type objects (Tholen, 1984). The X complex in the Tholen taxonomy comprises 
the E, M and P classes which
have very different inferred mineralogies but which are spectrally similar to
each other, with
featureless spectra in visible wavelengths.\\
The data were obtained during several observing runs 
in the 2004-2007 years at the NTT, TNG and IRTF telescopes. 
Sixteen asteroids were 
observed in the visible and near-infrared wavelength range, seven objects 
in the visible 
wavelength range only, and one object in the near-infrared wavelength range only.
We find a large variety of near-infrared spectral behaviors within the X class, and we 
identify  
weak absorption bands in spectra of 11 asteroids. 
Our spectra, together with  albedos published by 
Tedesco et al. (2002),  can be used to suggest new Tholen classifications for these objects.  We describe 1 A-type (1122), 1 D-type (1328), 1 E-type (possibly, 3447 Burckhalter), 10 M-types (77, 92, 184, 337, 417, 741, 758, 1124, 1146 and 1355), 5 P-types (275, 463, 522, 909, 1902), and 6 C-types (50, 220, 223, 283, 517, and 536). 
In order to constrain the possible composition of these asteroids,
we perform a least-squares search through the RELAB spectral
database.  Many of the 
best fits are consistent with meteorite analogue materials suggested in the published
literature.  In fact, we find that 7 of the new M-types can be fit with
metallic iron (or pallasite) materials, and that the low albedo C/P-type asteroids are best fitted with CM meteorites, some of which have been subjected to heating episodes or laser irradiation.  
Our method of searching for meteorite analogues emphasizes
the spectral characteristics of brightness and shape, and de-emphasizes
minor absorption bands. Indeed, faint absorption features like the 0.9 $\mu$m band seen on 4 newly classified 
M-type asteroids are not reproduced by the iron meteorites. In these cases,
we have searched for 
geographical mixture models that can fit the asteroid spectrum, minor bands, 
and 
albedo. We find that a few percent (less than 3\%) of orthopyroxene added to iron or pallasite meteorite, results in good spectral matches,  
reproducing the weak spectral feature around 0.9 $\mu$m seen on 92 Undina, 417 Suevia, and 1124 Stroobantia. For 337 Devosa, a mixture model that better reproduces 
its spectral behavior and the 0.9 $\mu$m feature is made with 
Esquel pallasite enriched with goethite (2\%).

Finally, we consider  the sample of the X-type asteroids we have when we
combine the present observations with previously published observations
for a total of 72 bodies. 
This sample includes M and E-type asteroid data
presented in Fornasier et al. (2008, 
2010). We find that the mean visible spectral slopes for the different E, M and 
P Tholen classes are very similar, as expected. 
An analysis of the X-type asteroid distribution in the main belt is also reported, following both the Tholen and the Bus-DeMeo taxonomies (DeMeo et al. 2009).
\end{abstract}

\begin{keyword}
Asteroids, Surfaces  \sep Spectroscopy \sep Meteorites 
\end{keyword}
\end{frontmatter}

\newpage
\section{Introduction}

The X complex in the Tholen taxonomy comprises the E, M and P classes which
have very different inferred mineralogies but which are spectrally similar 
to each other and featureless in the visible wavelengths (Tholen, 1984).  
By convention X-types with measured geometric albedos are classified into Tholen E-type
(high albedo, $p_{v} > 0.3$), M-type (medium albedo, $0.1 < p_{v} < 0.3$), or P-type
(low albedo, $ p{v} < 0.1$) asteroids (Tholen\&Barucci, 1989). The X-type
asteroid indicates an "EMP" asteroid for which we do not have albedo
information.

The X-type asteroids are distributed throughout the main belt (Moth\'e-Diniz et al. 2003) but
tend to be concentrated around 3.0 AU, and in the Hungaria region (Warner et al. 2009). The X complex includes bodies with very different mineralogies. 
M asteroids may be composed of metals such as iron and 
nickel and may be the progenitors of differentiated iron-nickel
meteorites (Gaffey, 1976; Cloutis et al, 1990,
Gaffey et al. 1993).  Enstatite chondrites have also been proposed
as M-type asteroid analogs (Gaffey 1976; Vernazza et al. 2009; Fornasier et al. 2010; Ockert-Bell et al. 2010). 
A metallic iron interpretation requires that M asteroids are
from differentiated parent bodies that were heated to at least 2000 $^{\circ} C$ to
produce iron meteorites (Taylor, 1992). 

E-type asteroids have high albedo values and their surface  compositions 
are consistent with iron-free or iron-poor  silicates  such as enstatite,
forsterite or feldspar. E-types are thought to be the parent bodies of the
aubrite (enstatite achondrite) meteorites (Gaffey et al. 1989, Gaffey et al.,
1992, Zellner et al., 1977, Clark et al. 2004a, Fornasier et al. 2008).  
P-type asteroids have low albedo values, and featureless red spectra. 
P-types are located mainly in the outer region of the main belt (Dahlgren \& Lagerkvist 1995, 	
Dahlgren et al. 1997, Gil-Hutton \& Licandro 2010) and in the Jupiter Trojan
clouds (Fornasier et al. 2004, 2007a). They are probably primitive bodies, however
they have no clear meteorite analogues.  P-type asteroids
are presumed to be similar to carbonaceous chondrites, but with a higher organic content to explain the strong red spectral slopes (Gaffey et al. 1989, Vilas et al. 1994).

In the last 15 years, several near-infrared spectral surveys have been devoted to the study of the X group and in particular the M and E-type asteroids, revealing that these bodies 
show great diversity in the infrared, even within the same class. In addition, polarimetric measurements (Lupishko \&
Belskaya 1989) and radar observations (Ostro et al., 1991, 2000; Margot and Brown, 2003; Magri et al.,
2007; Shepard et al., 2008, 2010)  of selected M-type asteroids have revealed surface
properties that are in some cases inconsistent with metallic iron. 
Moreover, spectra of M-types are not uniformly featureless as 
initially believed; 
several spectral absorption bands have been detected in the visible and near
infrared ranges. Faint absorption bands near 0.9 and 1.9
$\mu$m have been identified on the surfaces of some M-types (Hardersen 
et al., 2005; Clark et al., 2004b; Ockert-Bell et al., 2008; Fornasier et al., 2010), and of some E-type asteroids (Clark et al. 2004a; Fornasier et al. 2008) and
attributed to orthopyroxene. A very peculiar absorption band, centered around
0.49-0.50 $\mu$m, extending from about 0.41 to 0.55 $\mu$m, was found on some E-type spectra (Burbine et al, 1998; Fornasier \& Lazzarin, 2001; Clark et al. 2004a,b;
Fornasier et al., 2007b, 2008), and attributed to the presence of sulfides such as
troilite or oldhamite (Burbine et al., 1998, 2002a). Busarev (1998) detected weak
features tentatively attributed to pyroxenes (at 0.51 $\mu$m) and oxidized or
aqueously altered mafic silicates (at 0.62 and 0.7 $\mu$m) in the spectra of 2
M-type asteroids. \\

Of note is the evidence for hydrated minerals on the surfaces of some  E and M-type asteroids (Jones et al, 1990; Rivkin et al, 1995,
2000, 2002), inferred from  the identification of an absorption feature around 3
$\mu$m.  This feature is typically attributed 
to the first overtone of H$_{2}$O and to OH vibrational
fundamentals in hydrated silicates. If the origin of the 3 $\mu$m band on E and 
M-type asteroids is actually due to hydrated minerals, then these objects 
are not all igneous as previously believed and the thermal scenario
in the inner main belt would need revision.
Gaffey et al. (2002) propose alternative explanations for the 3 $\mu$m band: materials normally considered to be anhydrous containing structural OH; troilite, an anhydrous mineral that shows a feature at 3 $\mu$m, 
or xenolithic hydrous meteorite components on asteroid surfaces from impacts 
and solar wind implanted H. Rivkin et al. (2002) provide arguments
refuting the Gaffey suggestions, and provide evidence in support of the
hydrated mineral interpretation of the observations.\\

The X complex is consistently identified in the two taxonomies 
recently developed based on CCD spectra acquired in the visible (Bus \& Binzel 2002) or in the visible and near infrared range (DeMeo et al., 2009), even though there is not a direct one-to-one correspondence between the Bus and Tholen classes. For instance 92 Undina (Tholen M-type), and 65 Cybele (Tholen P-type) are both  classified as Xc-type in the Bus taxonomy. A description of the X complex classes (X, Xk, Xe, Xc) in the Bus \& Binzel (2002) classification is summarized in Clark et al. (2004b), while in section 6 we describe the X complex classes in the Bus-DeMeo taxonomy.

Aiming to constrain and understand the composition of the
asteroids belonging to the mysterious X group, we have carried out a
spectroscopic survey of these bodies in the visible and near infrared range at
the 3.5m Telescopio Nazionale Galileo (TNG), in La Palma, Spain, at the 3.5m
New Technology Telescope (NTT) of the European Southern 
Observatory, in La Silla,
Chile, and at the Mauna Kea Observatory  3.0 m NASA Infrared Telescope Facility
(IRTF) in Hawaii, USA. 
Results on asteroids classified as E and M-type following the Tholen 
taxonomy are presented elsewhere (Fornasier et al. 2008, 2010).
In this paper we present new VIS-NIR spectra of 24 asteroids belonging to
the X type as defined by Tholen \& Barucci (1989), that is an "E--M--P" type
asteroid for which albedo information was not available at the time of their
classification. To constrain their compositions, we conduct a search for
meteorite analogues using the RELAB database, and we model the asteroid
surface composition with geographical mixtures of selected minerals when a
meteorite match is not satisfactory. 
In addition, we present an analysis of X complex spectral 
slope values and class distributions in the asteroid main belt, 
where we include previously published observations of M and E-type
asteroids obtained during the same survey (Fornasier et al. 2008, 2010).

\section{Observations and Data Reduction}

[HERE TABLE 1 ]

The data presented in this work were mainly obtained  during 2 runs 
(February and November 2004) at the Italian Telescopio Nazionale Galileo 
TNG of the European Northern Observatory (ENO) in La Palma, Spain, and 2 runs (August 2005, and January 2007) at the New Technology Telescope (NTT)
of the European Southern Observatory (ESO), in Chile. Two asteroids (92 Undina, 275 Sapientia),
investigated in the visible range at the NTT and TNG telescopes, were separately
observed in the near infrared at the Mauna Kea Observatory  3.0 m NASA Infrared
Telescope Facility (IRTF) in Hawaii during 2 runs in July 2004 and September 2005 (Table~\ref{tab1}).
We had a total of 11 observing nights. Here
we present the results of our observations of 24 X-type asteroids: for 16
objects we obtained new visible and near-infrared wavelength spectra (VIS-NIR),
for one object we obtained only a near-infrared spectrum (NIR), 
and for 7 objects we obtained only visible range spectra (VIS).

\begin{figure*}[b]
\includegraphics[width=14cm]{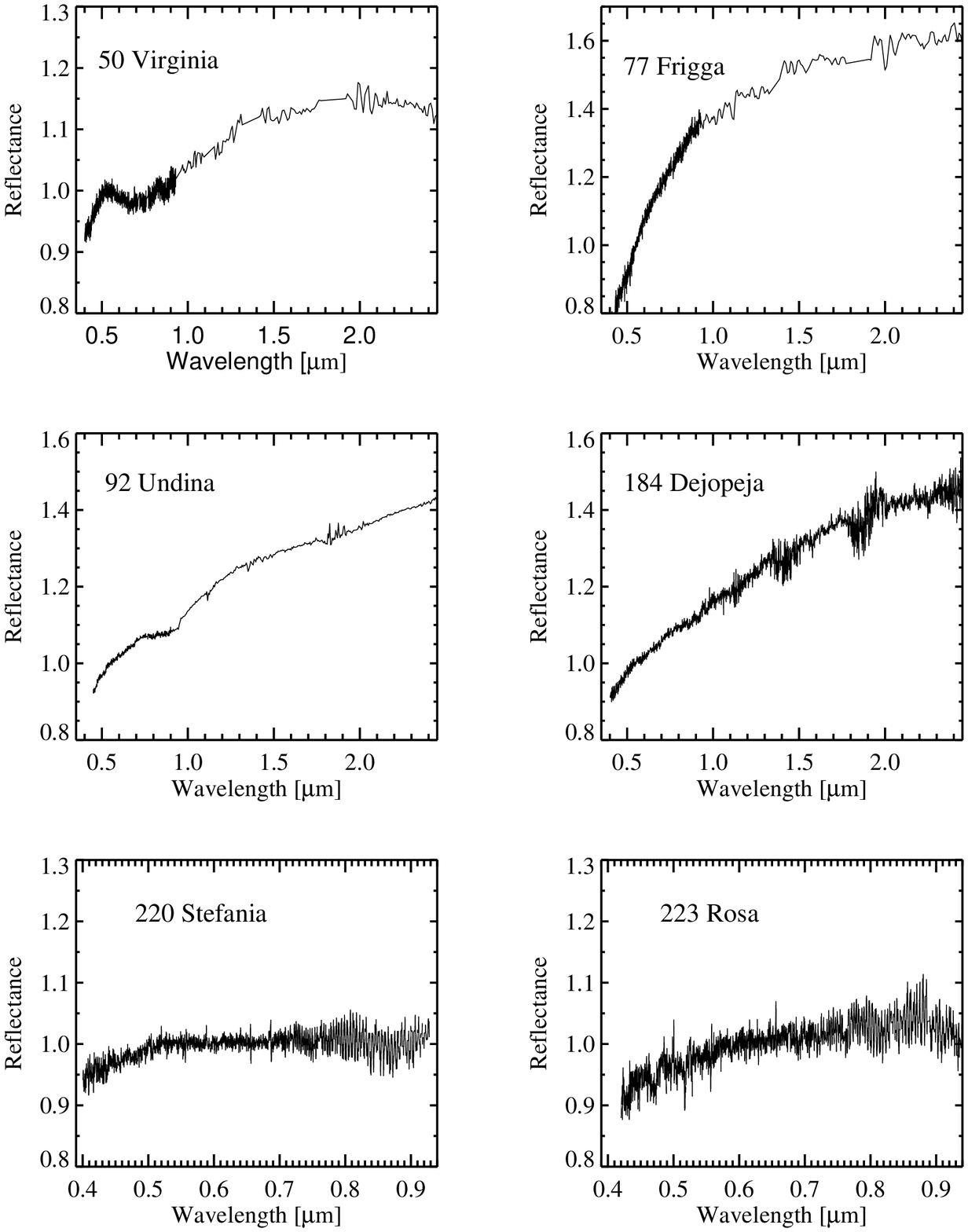}
\caption{}
\label{fig1}
\end{figure*}

\begin{figure*}
\includegraphics[width=14cm]{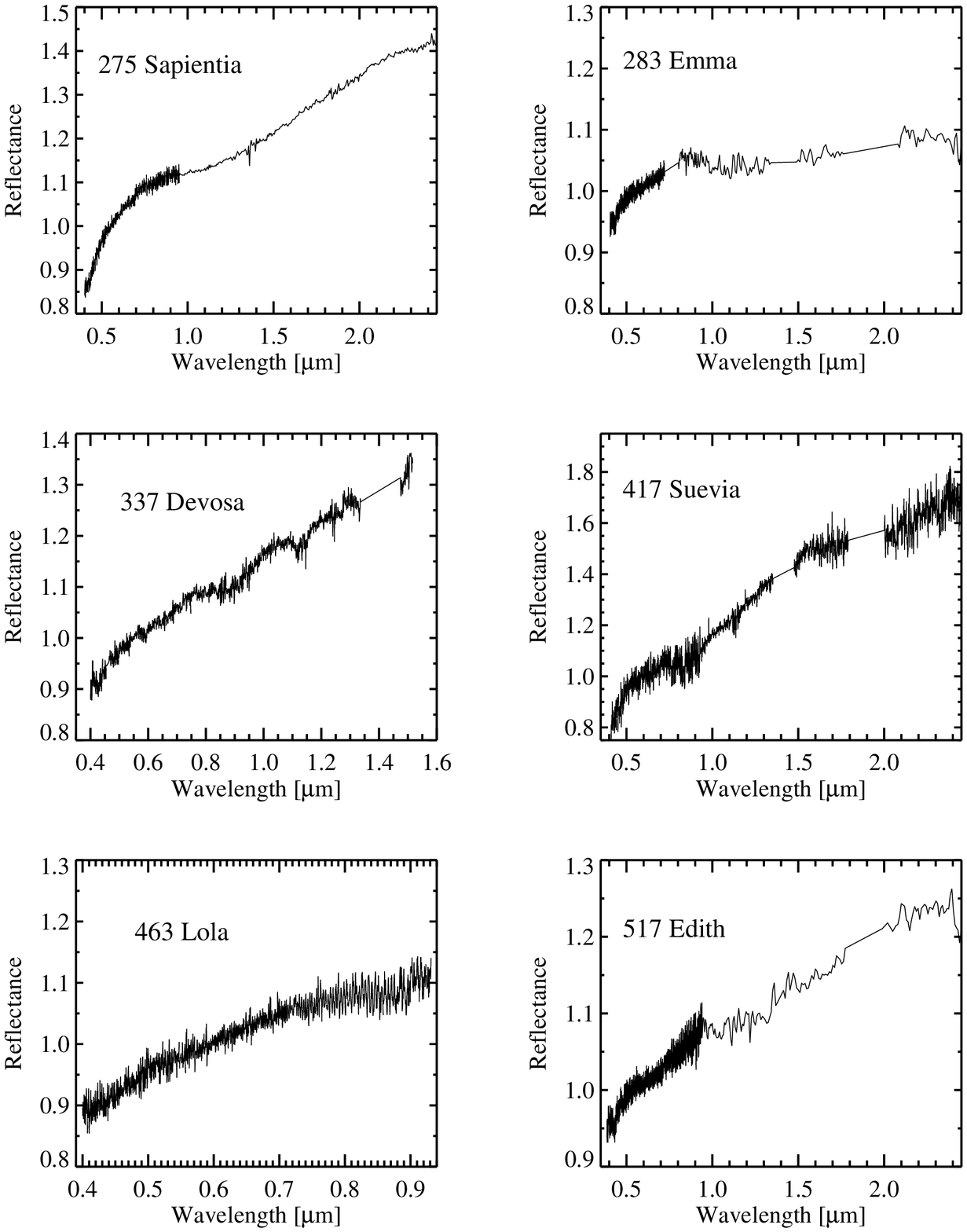}
\caption{}
\label{fig2}
\end{figure*}

\begin{figure*}
\includegraphics[width=14cm]{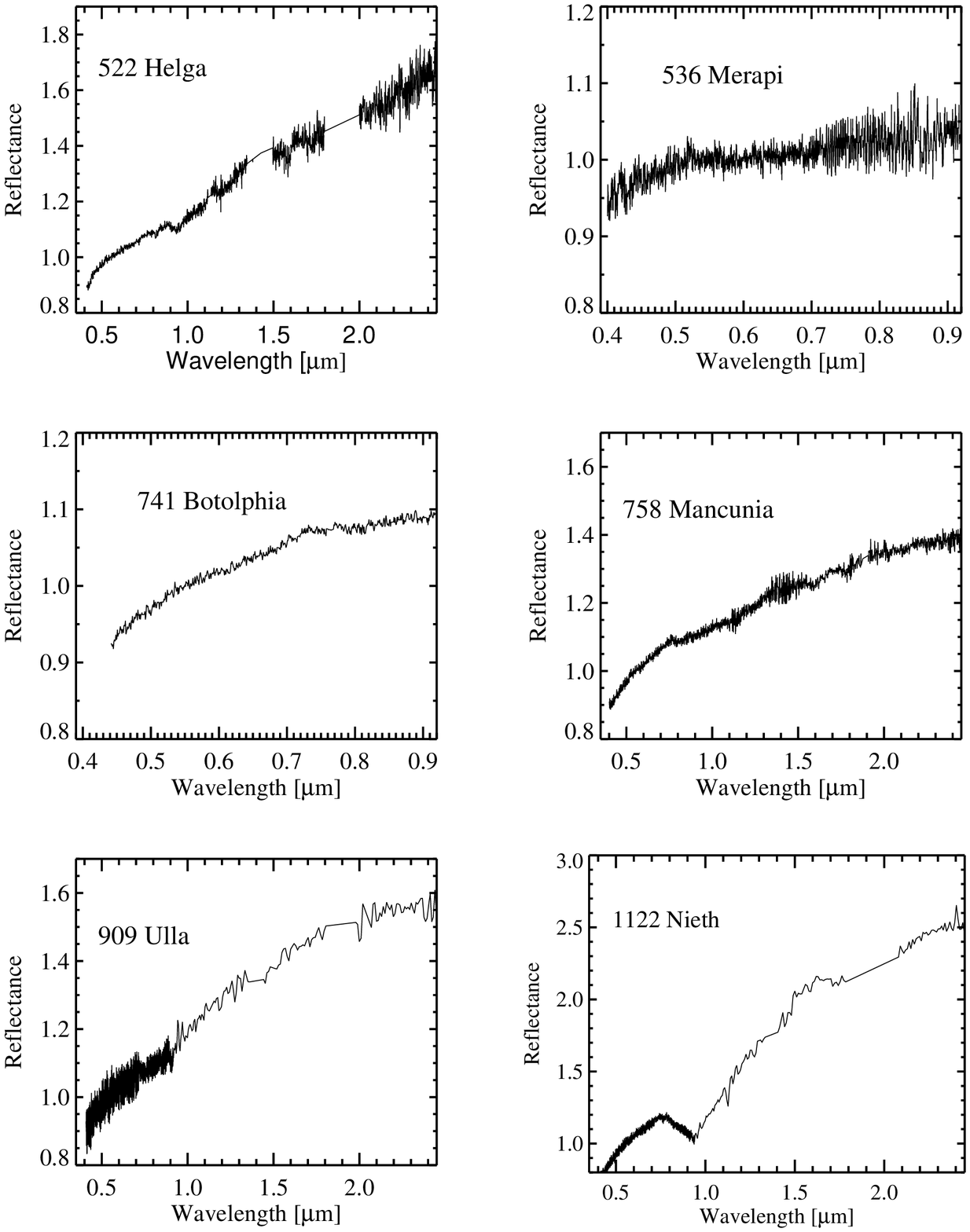}
\caption{}
\label{fig3}
\end{figure*}

The instrument setups are the same as used 
in Fornasier et al. (2008, 2010). We refer the reader to those papers
for complete
descriptions of observations and data reduction techniques.
The spectra of the observed asteroids, all normalized at 0.55 $\mu$m, are shown
in Figs.~\ref{fig1}-- \ref{fig4}, and the observational conditions are reported in Table~\ref{tab1}.

[Here Figs 1,2,3, and 4]

To analyze the data, spectral slope values were calculated with linear
fits to different wavelength regions: $S_{cont}$ is the spectral slope in the whole
range observed for each asteroid, $S_{UV}$  in the 0.49-0.55 $\mu$m range, $S_{VIS}$ is the slope in the 0.55-0.80 $\mu$m
range, S$_{NIR1}$ is the slope in the 1.1-1.6 $\mu$m range, and S$_{NIR2}$ is
the slope in the 1.7-2.4 $\mu$m range. Values are reported in
Table~\ref{slope}. 
Band centers and depths were calculated for each asteroid showing an 
absorption feature, following the Gaffey et al. (1993) method. 
First, a linear continuum was fitted at the edges of the band, 
that is at the points on the spectrum outside the absorption 
band being characterized. Then the asteroid spectrum was divided 
by the linear continuum and the region of the band was fitted with a 
polynomial of order 2 or more. The band center was then calculated 
as the position where the minimum of the spectral reflectance 
curve occurs, and the band depth as the minimum in the ratio of the 
spectral reflectance curve to the fitted continuum (see Table~\ref{band}).

[HERE TABLE 2]

[HERE TABLE 3]

\section{Spectral Analysis and Absorption Features}

The observed asteroids show very different spectral behaviors
(Figs~\ref{fig1}-- \ref{fig4}). Eleven asteroids show at least one absorption
feature (Table~\ref{band}). 

50 Virginia (see Fig. 1 and Table 3) shows spectral features centered at 0.43, 0.69, and 0.87 $\mu$m (Table~\ref{band}) associated with aqueous alteration products (Vilas et al. 1994, Fornasier et al. 1999), and seems to be a typical 'hydrated' C-type asteroid. 

Five asteroids (50 Virginia, 283 Emma, 337 Devosa, 517 Edith, and 1355 Magoeba) show a 
weak band at 0.43 $\mu$m. For 50 Virginia, 283 Emma, and 517 Edith, 
the band might be associated with an Fe$^{3+}$  spin-forbidden 
transition in the iron sulfate jarosite, as suggested by  
Vilas et al. (1993) for low-albedo asteroids. For 337 Devosa and 1355 Magoeba, 
the band might be associated with chlorites and Mg-rich serpentines,  
as suggested by King \& Clark (1989) for enstatite chondrites, 
or to clinopyroxenes such us pigeonite or augite as
suggested by Busarev (1998) for M-asteroids. 

The 1355 Magoeba spectrum also shows a band 
centered at $\sim$ 0.49 $\mu$m that
resembles the absorption seen on some E-type (subclass EII) asteroids (Fornasier \& Lazzarin 2001; Fornasier et al. 2007b; 2008, Clark et al. 2004a), where it is attributed to sulfides such
as oldhamite and/or troilite (Burbine et al. 1998, 2002a, 2002b). Nevertheless, the estimated albedo of 1355 Magoeba has a moderate value (0.267$\pm$0.095, Gil-Hutton et al. 2007) more consistent with a Tholen M-type classification.

92 Undina has a very weak band centered at $\sim$ 0.51 $\mu$m that is similar to the Fe$^{2+}$
spin-forbidden crystal field transitions in terrestrial and lunar pyroxenes (Burns et
al. 1973, Adams 1975, Hazen et al. 1978). This band has been previously detected
by Busarev (1998) in the spectra of two M asteroids, 75 Eurydike and 201
Penelope. 

An absorption feature has been identified  in the 0.9 $\mu$m
region in the spectra of 4 X-type asteroids that
have albedos placing them in the Tholen M-class (92 Undina,
337 Devosa, 417 Suevia, and 1124 Stroobantia). 
The band center ranges from 0.86 $\mu$m to 0.92 $\mu$m with a
band depth (as compared to the continuum) of 3--6 \%.
Also 522 Helga, having a low albedo (4\%)
consistent with the Tholen P-type classification, exhibits a faint band centered at 0.94 $\mu$m.  
This band was previously 
reported by several authors in the spectra of Tholen M, E and X-type
asteroids (Hardersen et al. 2005; Clark et al. 2004a, 2004b, Fornasier
et al. 2008, 2010) and is attributed to low-Fe, low-Ca orthopyroxene
minerals. 

A peculiar feature centered at 1.6 $\mu$m was found on the spectrum of 758 Mancunia. This band resembles one seen on 755 Quintilia (Fornasier et al., 2010, Feiber-Beyer et al. 2006) whose interpretation is still unknown. No bands due to silicates in the 0.9 and 1.9 $\mu$m regions have been identified in our spectra, except for a change in slope at around 0.77 $\mu$m.

Finally, 1122 Nieth shows a strong 0.96 $\mu$m absorption band (depth of 24\%) and a steep slope in the NIR region, typical of olivine-rich bodies. \\

Several asteroids of this survey were observed previously by other authors.
Five X-type asteroids (50 Virginia, 283 Emma, 337 Devosa, 517 Edith, and 909 Ulla) of the Clark et al. (2004b) survey of the X complex were 
also studied in this work. The spectral behavior for these objects looks quite similar,
except for 283 Emma (our spectrum is flat in the NIR, while the Clark et
al. spectrum has a concave shape with a higher spectral gradient),  and 909 Ulla
(our spectrum has a higher spectral gradient 
as compared to the Birlan et al. (2007) and the Clark et al. (2004b) spectra).
Ockert-Bell et al. (2008, 2010) presented near-infrared spectra of 77 Frigga and 758 Mancunia,
that look slightly different compared to ours. In particular, they detect faint spectral bands in the 0.9 and 1.9 $\mu$m region for 758 Mancunia, and in the 0.9 $\mu$m region for 77 Frigga, bands that are not detected in our spectra.
1355 Magoeba and 3447 Burckhalter were observed in the visible
range by Carvano et al. (2001), and their data look very similar to ours.\\ 
In sum, comparing our spectra with those in the existing literature, we suggest
that asteroids 77 Frigga, 283 Emma, 909 Ulla, and 758 Mancunia may display surface variability as they show different spectral behaviors throughout independent observations. For 758 Mancunia this variability is confirmed also by radar measurements that show the radar cross-section and the polarization ratio to
vary considerably with rotation phase (Shepard et al. 2008). For the other 3 asteroids, we cannot exclude that some spectral differences may be linked with unresolved observational differences (such as
background stars, or undetected troubles in removing the atmosphere).

\section{Tholen Taxonomic Classification}

Our targets were classified as belonging to the Tholen X class on the basis of
their spectra from 0.4 to 1 $\mu$m and because their albedos were not known
at the time of classification.
Since the original X-type classification in the Tholen taxonomy, 
the albedo value has become available for most of the asteroids 
we observed (Tedesco et al., 2002). Taking into account this important 
information, together with the visible and near infrared (when available) 
spectral behavior,  we suggest a re-classification of the X-type 
asteroids in our sample following the Tholen-classification scheme. 
The suggested new classifications are reported in Table ~\ref{slope}.

\begin{figure*}
\includegraphics[width=14cm]{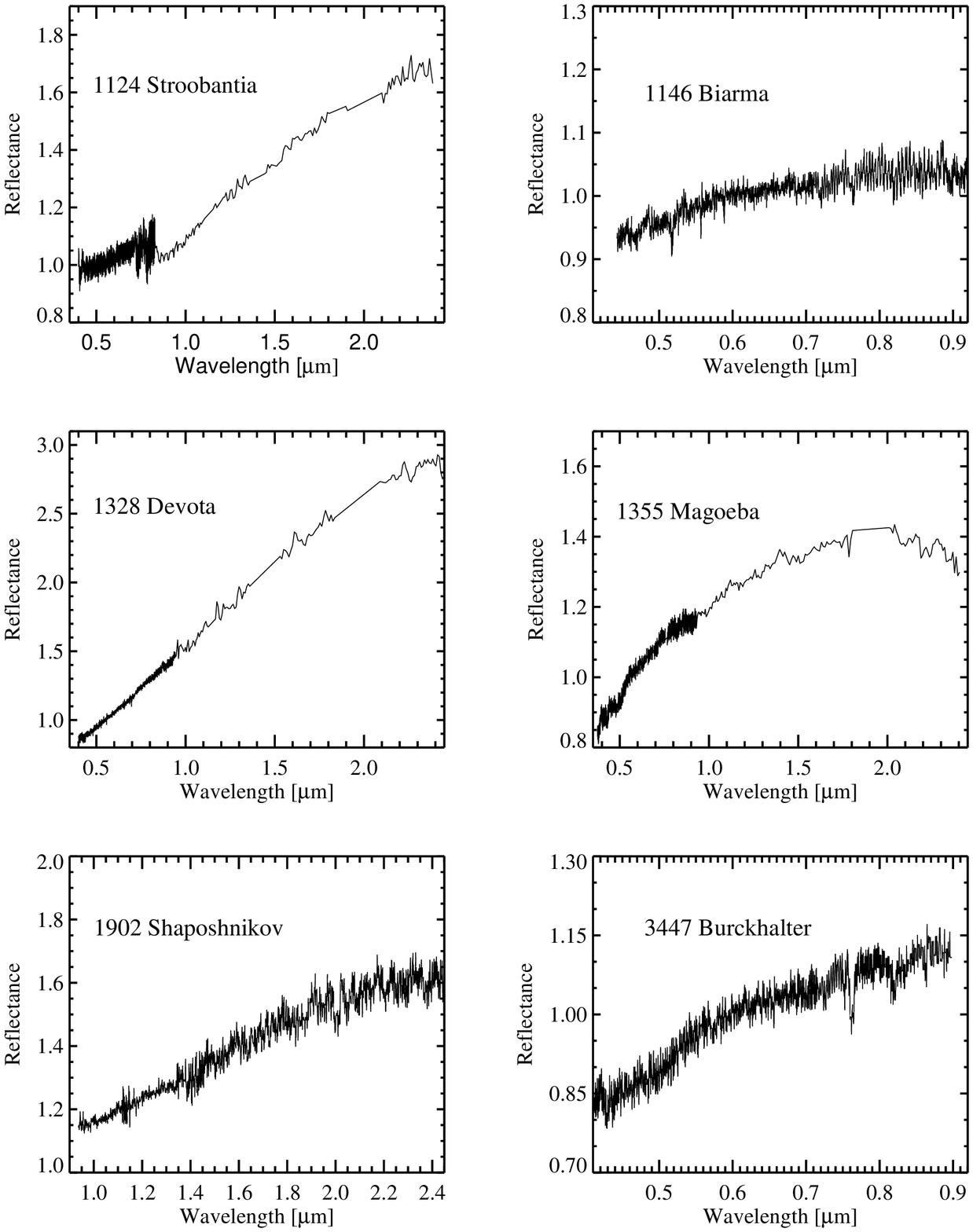}
\caption{}
\label{fig4}
\end{figure*}

1122 Nieth (Fig.~\ref{fig4}), initially classified as X-type in the Tholen taxonomy,
has atypical spectral properties that diverge from all other asteroids.
The near infrared spectrum of Nieth
clearly shows a strong 0.96 $\mu$m absorption band 
and a steep slope in the NIR
region. We infer that this asteroid belongs 
to the olivine-rich A-types (used in
the Tholen, Bus, and Bus-DeMeo taxonomic systems).

1328 Devota has a very steep featureless red spectrum. Considering its low albedo value (4\%), 
Devota falls within the D class in the Tholen taxonomy (Fig.~\ref{fig4}).

We suggest that the ten asteroids (77 Frigga, 92 Undina, 
184 Dejopeja, 337 Devosa, 417 Suevia, 741 Botolphia, 
758 Mancunia, 1124 Stroobantia, 1146 Biarmia, and 1355 Magoeba) with 
moderate albedo values (0.1--0.3) are M-type asteroids while those with albedos lower than 10\% fall in the C/P classes. 
To discriminate between C and P-type asteroids, we 
analyzed spectral slope values in the visible 
range ($S_{VIS}$ from 0.55 to 0.80 $\mu$m) and searched for 
the UV drop-off below $\sim 0.5 ~\mu$m that is
typical of C-type asteroids. Comparing the $S_{VIS}$ slopes 
with the $S_{UV}$ slopes (calculated from 0.49 to 0.55 $\mu$m) 
of the low albedo asteroids, we find two distinct distributions 
correlated with the C and P-classes. The 6 asteroids 
(50, 220, 223, 283, 536, and 517) having the UV absorption and
$S_{vis} < 2 \%/10^{3}$\AA\ belong to the C-type, while the 
4 asteroids (275, 463, 522, and 909) having 
$3 <S_{vis} < 6 \%/10^{3}$\AA\ belong to the P-type. 
1902 Shaposhnikov, observed only in the near infrared 
range, is probably also a P-type, while 3447 Burckhalter, for which the albedo value is 0.336$\pm$0.164 
 (Gil-Hutton et al. 2007) could possibly be an E-type.

\section{Meteorite Spectral Matches and Mixing Models}

\subsection{Meteorite Analogues}

To constrain the possible mineralogies of our asteroids, we conducted  
a search for meteorite and/or mineral spectral matches.
We sought matches only for objects observed across the entire 
VIS-NIR wavelength range.  A complete description of 
the search methodology can be found in Clark et al. (2010) and
Fornasier et al. (2010).  We used the  
publicly available RELAB spectrum library (Pieters 1983), which  
consisted of nearly 15,000 spectra in November of 2008.  RELAB spectra were 
normalized to 1.0 at 0.55 $\mu$m, and then a Chi-squared value 
was calculated for each RELAB spectrum relative to  
the normalized input asteroid spectrum. The RELAB spectra were sorted according  
to Chi-squared values, and then visually examined for dynamic weighting of  
spectral features by the spectroscopist.  
Given similar Chi-squared  
values, a match that mimicked spectral features was preferred over a  
match that did not.  We visually examined the top ~200 Chi-squared  
matches for each asteroid and we constrain the search for 
analogues to those laboratory spectra with
brightness (reflectance at 0.55 $\mu$m) roughly comparable to the albedo of
the asteroid.  
It must be noted that the asteroid data are disk integrated, 
which could mask spectral variations due to differences in 
particle size, mineralogy, or abundance. However the meteorite 
and mineral spectra are well characterized powders measured under 
well defined laboratory conditions. It should also be noted that
 an asteroid's albedo 
is defined at phase angle zero,
whereas meteorites' reflectance is taken at phase angle  $> 5^{o}$.  
For the asteroid albedo value, we used the IRAS  
albedo published by Tedesco et al. (2002).  For the darker asteroids  
(below 10\% albedo), this is a fairly well-constrained value.  The  
uncertainty for brighter asteroid albedos (above 10\% albedo) is  
relatively larger. To account for this, we filtered out lab spectra if
brightness differed by more than $\pm$ 3\% in absolute albedo for the darker  
asteroids (albedos less than 10\%), and we filtered out lab spectra if  
brightness differed by more than $\pm$ 5\% in absolute albedo for the  
brighter asteroids  (albedos greater than 10\%).  For example, if the asteroid' s albedo was 10\%, we searched over all lab spectra with  
0.55 $\mu$m reflectance between 5\% and 15\%.  Once the brightness filter was  
applied, all materials were normalized before comparison by least- 
squares. 

Our search techniques effectively emphasized the  
spectral characteristics of brightness and shape, and de-emphasized  
minor absorption bands and other parametric characteristics.  As such,  
we suggest that our methods are complementary to band parameter  
studies.

[HERE TABLE 4]

[HERE FIGURES 5 and 6]

The best matches between the observed X-type asteroids and meteorites
from the RELAB database are reported
in Table~\ref{tab}. We matched only the 15 asteroids 
observed both in the visible and near infrared range that
have published albedo values. Some meteorite matches are quite good, while for some asteroids the best meteorite 
match we found does not satisfactorily reproduce both the visual and near infrared
spectral behavior. In particular our attempt to find a meteorite or mineral match failed for the 3 asteroids 1122 Nieth (A-type), 1328 Devota (D-type), and 1355 Magoeba (M-type).

\begin{figure*}
\includegraphics[width=14cm]{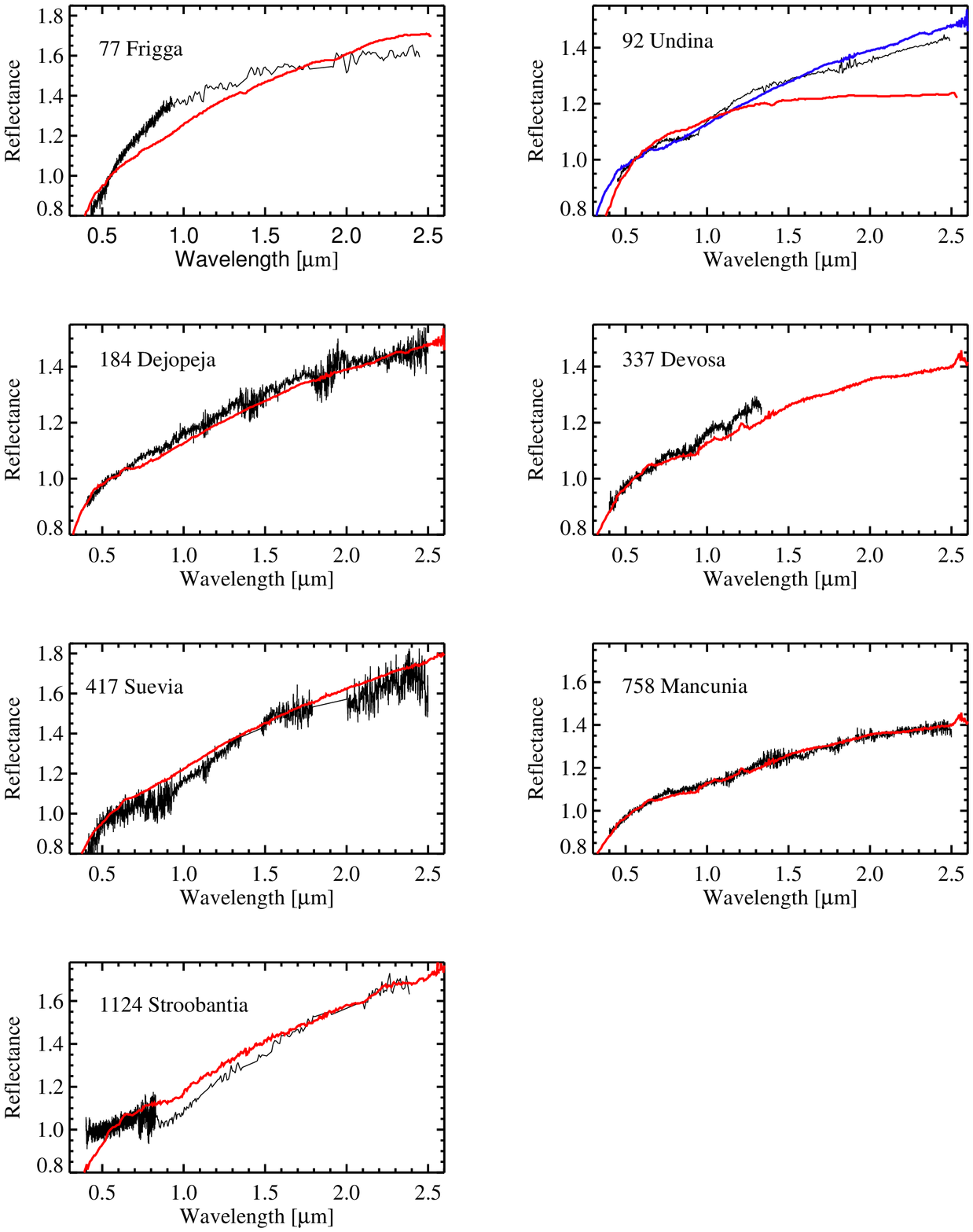}
 \caption{}
   \label{fig5met}
 \end{figure*}

The best matches we found 
for 7 moderate albedo M-type asteroids are presented in Figure~\ref{fig5met}.  Two asteroids, 
741 Botolphia and 1146 Biarmia were not compared with RELAB spectra
due to the lack of near-infrared spectral data. 
We find that iron or pallasite meteorites are the best matches both for the featureless M-type asteroids and for those having minor absorption bands. This result supports the link between M-type asteroids and iron or pallasite meteorites suggested in the literature by several authors. In this small sample, no enstatite chondrites, which have also been suggested to be linked with M-type asteroids, have been found as the best meteorite analogue.

The strongest analogy 
we found is between 758 Mancunia and a sample of the iron meteorite Landes which has a large number of silicate inclusions in a nickel-iron matrix.
It must be noted that no specific grain-size sampling
filter was applied, and of concern is the fact that Mancunia's best match is with a spectrum obtained of the cut
slab surface of the Landes iron meteorite. 

758 Mancunia has a radar albedo of 0.55$\pm$0.14 
(Shepard et al. 2008) one of the highest values measured for the 
asteroids, suggesting a very high metal content. Shepard et al. (2008) also found variations in the radar cross-section and polarization ratio with Mancunia's rotational phase, variations that are not related to shape but to the regolith depth, porosity, or near-surface roughness. The Ockert-Bell et al. (2010) Mancunia spectrum shows absorption bands in the 0.9 and 1.9 $\mu$m region, and is found to be similar to the ordinary chondrite Paragould. Our spectrum does not show these features, but a change of slope at 0.77 $\mu$m and a faint (1.3\% depth) feature centered at 1.6 $\mu$m, similar to that observed on 755 Quintilia (Fornasier et al. 2010, Fieber-Beyer et al. 2006), whose origin is not yet understood.  Considering both radar and spectral observations, 758 Mancunia could present an heterogeneous surface composition, with a high metal content but also the presence of some silicates that are not uniformly distributed on the surface.

The Landes iron meteorite sample was found to be the best spectral match also for 337 Devosa, even if the meteorite does not fully reproduced the 0.88 $\mu$m absorption band of the asteroid. A similar problem occurred for 417 Suevia and 1124 Stroobantia, where different grain sized samples of the iron meteorite DRP78007 are found to be the best spectral matches, but they did not reproduce the asteroids absorption bands in the 0.9 $\mu$m region. 
We note that grain size effects can play an important role in the spectral behavior of iron-nickel meteorites. In fact Britt and Pieters (1988) found that
the spectra of M-type asteroids show good agreement with those of iron
meteorites with surface features in the range of 10 $\mu$m to 1 mm, that is
larger that the wavelength of incident light.  Meteorites with these roughness
values are diffuse reflectors and show the classic red slope continuum of iron,
with practically no geometric dependence on reflection. On the other hand, a  
decreasing of the meteorite surface roughness changes the reflectance
characteristics:  complex scattering behavior is seen for roughness in the
0.7-10 $\mu$m range, while for roughness values $<$ 0.7 $\mu$m the reflectance
is characterized by two distinct components, the specular one which is bright and
red sloped, and the nonspecular one which is dark and flat.  \\

For 77 Frigga we were not able to find any meteoritic/mineral 
analog within the medium albedo range that could
reproduce Frigga's red visible slope and flat near-infrared slope. 
The best spectral match is shown in Fig.~\ref{fig5met}.
The iron meteorite Chulafinnee
models the spectral behavior below 0.55 $\mu$m very well,
and the albedo values are similar, 
however the change in spectral slope around 0.9 $\mu$m is not
reproduced.

For 92 Undina we propose two different meteorites: 
the best spectral match is with a metal-rich powder sample of the pallasite meteorite 
Esquel, whose albedo of 0.14 is lower than the 0.25 value
of the asteroid. The iron meteorite Babb's Mill
was a better numerical fit (minimum chi squared fit)
and had a more comparable albedo, but was not favored because 
Babb's Mill has a near  infrared spectrum that is 
much flatter than that of Undina.

\begin{figure*}
\includegraphics[width=14cm]{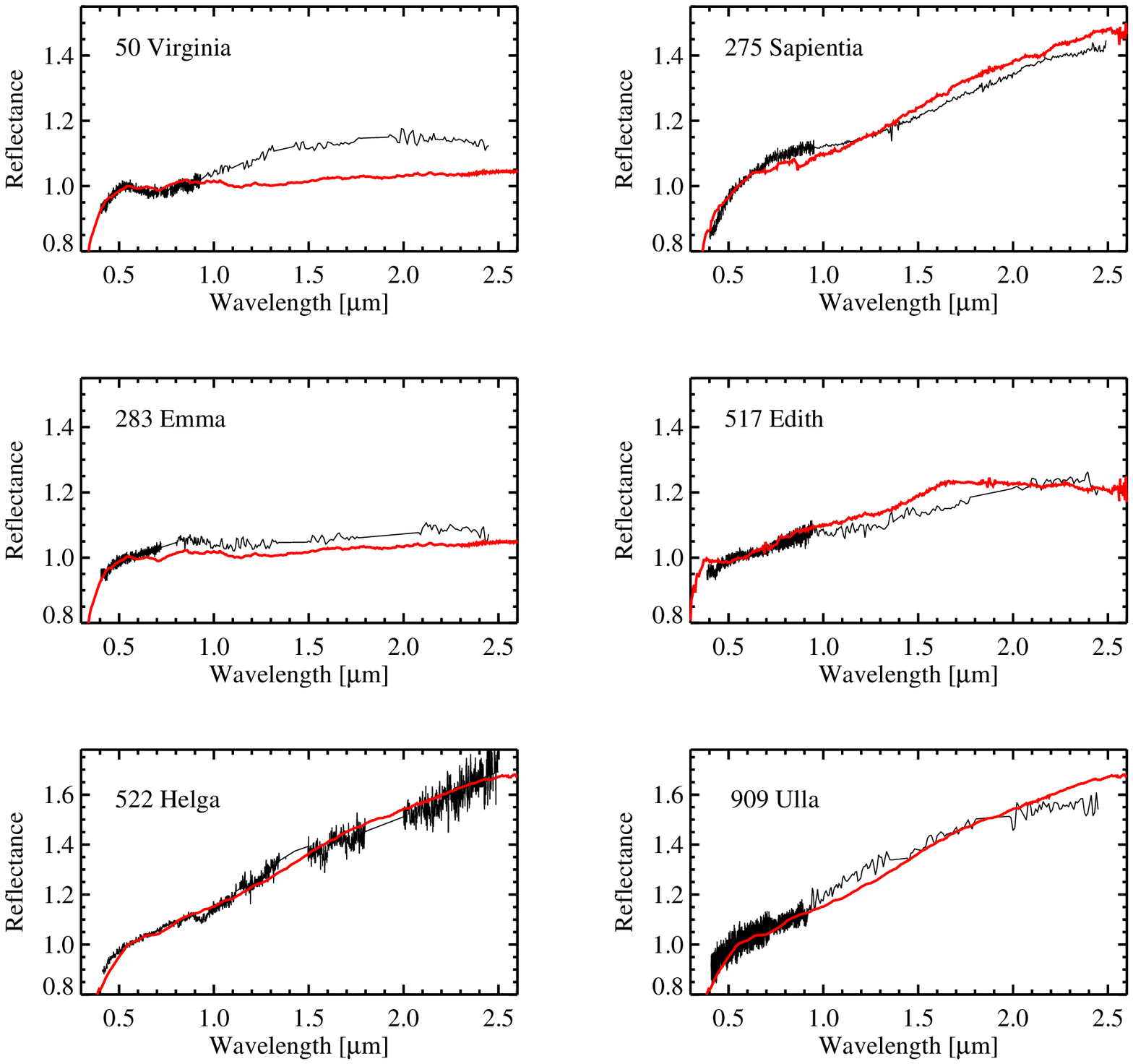}
 \caption{}
   \label{fig6met}
 \end{figure*}

In Figure~\ref{fig6met} we present the best matches 
proposed for six low albedo asteroids. 
It has long been noticed that reflectance spectra of carbonaceous
chondrites are similar to those of the low-albedo asteroids
(Gaffey \& McCord 1978; Hiroi et al. 1996). In particular, CM meteorites 
show features due to aqueous alteration processes and are linked
to low albedo aqueous altered asteroids like the C and G-types (see Vilas et al. 1994, Burbine 1998, Fornasier et al. 1999).   
Indeed, all the observed low albedo asteroids are best fit with CM meteorites, either unaltered (50 Virginia and 283 Emma) 
or altered by heating episodes (275 Sapientia and 517 Edith) 
or laser irradiation (522 Helga and 909 Ulla) (see Fig.~\ref{fig6met}). 
The CM2 MET00639 meteorite matches very well the visible range spectrum of 50 Virginia and in particular the 0.7 $\mu$m band attributed to $Fe^{2+}\rightarrow Fe^{3+}$ charge transfer absorptions in phyllosilicate minerals. Nevertheless, the meteorite does not well reproduce the asteroid's near-infrared spectrum. The same meteorite was found as a best match for the near infrared region of 283 Emma, but it does not well reproduce the spectrum in the visible range, where the asteroid is featureless. The absence of the 0.7 $\mu$m band in the asteroid spectrum led us to discard this as a spectral match.

The P-type asteroids 275 Sapientia and the C-type 517 Edith have as best match two different samples of the Murchison carbonaceous chondrite heated at 600$^{o}$C (grain size $<$ 63 $\mu$m) and 700$^{o}$C (63 $\mu$m $<$  grain size $<$ 125 $\mu$m), respectively. It must be noted that at these temperatures the phyllosilicates are dehydrated and transformed into olivine and pyroxene and that the 3 $\mu$m hydratation band vanishes (Hiroi et al., 1996).  
Our spectral match may indicate that Sapientia and Edith have no hydrated silicates on their surfaces. If hydrated silicates were originally present, the asteroids may have experienced important thermal episodes that dehydrated them.

The P-type asteroids 522 Helga and 909 Ulla are reproduced by a sample of the CM meteorite Migei after laser irradiation. This meteorite is composed of a black matrix with olivine-rich chondrules, olivine aggregates and individual grains, carbonates, and sulfides (Moroz et al. 2004). Its unaltered spectrum is dark and shows absorption features related to hydrated silicates such as the 2.7-3 $\mu$m band, the UV-falloff and a 0.75 $\mu$m band. Once irradiated, the meteorite is dehydrated, the absorption bands are significantly weakened (Moroz et al. 2004), and the material remains dark -- most probably due to abundant submicron inclusions of Fe-rich phases finely dispersed in the glassy mesostasis (Shingareva et al. 2004). 
Because laser irradiation is a laboratory technique for simulating micrometeorite bombardment, the spectral match between Helga and Ulla and a laser irradiated sample of CM Migei may indicate that the surfaces of these outer belt asteroids, if composed of CM carbonaceous-like materials, may be spectrally reddish due to micrometeoritic bombardment.

\subsection{Mixing Models}
[HERE FIGURE 7]

[HERE TABLE 5]

To constrain the surface compositions of the investigated 
X-type asteroids and the materials needed to reproduce the weak 0.9 $\mu$m band seen on some spectra
we considered geographical (spatially segregated) mixtures of 
several terrestrial and meteoritic materials in different grain 
sizes: in particular we took into account all the samples 
included in the US Geological Digital Spectral Library 
($http://speclab.cr.usgs.gov/spectral-lib.html$), in 
the RELAB database, and in the ASTER spectral library ($http://speclib.jpl.nasa.gov$),
together with 
organics solids (e.g. kerogens by Clark et al. 1993 and Khare et al. 1991;
Titan tholins from Khare et al. 1984; and
Triton tholins from McDonald et al. 1994) and 
amorphous carbon (by Zubko et al. 1996).
The synthetic spectra  were compared with 
the asteroid spectra, using the known IRAS albedo, and 
the VIS-NIR spectral behavior and continuum slopes. 
We present these matches as non-unique examples of how
geographical mixtures can be used to explain some of the
variety found in X-type asteroid spectra.  We do not intend
to indicate that these are unique derivations of the composition
of these asteroids.  Such an inversion requires much more
information than we currently possess (grain sizes, optical
constants, endmembers present, etc.).
Nevertheless, the models we present can
provide a first order understanding of
 possible surface composition.

\begin{figure*}
\includegraphics[width=14cm]{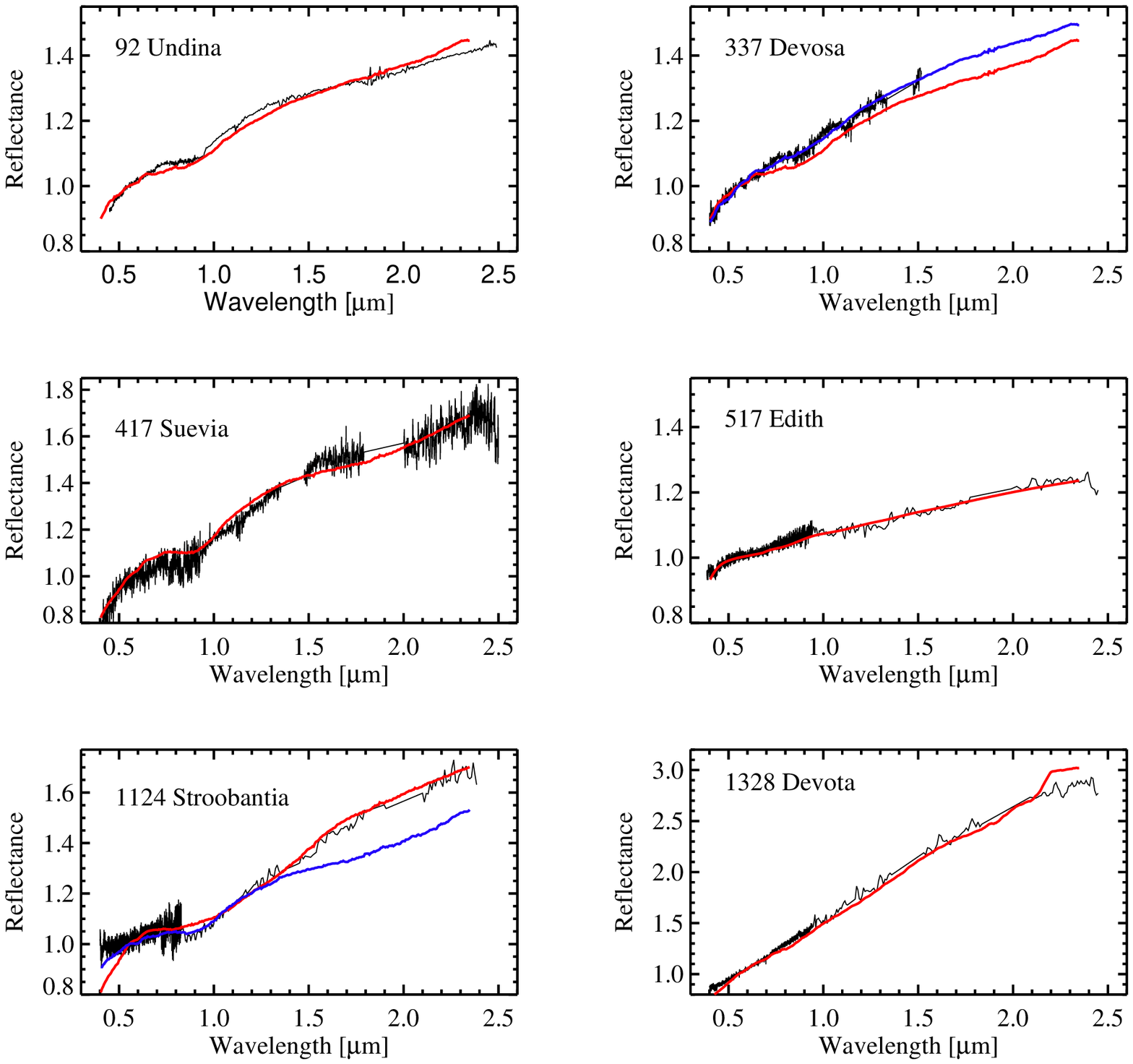}
 \caption{}
   \label{modelli}
 \end{figure*}

In Figure~\ref{modelli} and Table~\ref{models} we present 
mixing models 
for the 4 X-type asteroids showing the 0.9 $\mu$m band (this band is
attributed to low-Fe, 
low-Ca orthopyroxene and is not reproduced by the meteorite analogues proposed 
in Table~\ref{tab}). We also present
the low albedo asteroids 517 Edith (for which 
the spectral match with a heated CM 
meteorite is poor in the 1--2 $\mu$m region), and 1328 Devota (for which we 
cannot find any satisfactory meteorite analogue).

As shown in Fig.~\ref{fig5met} and discussed in Section 5.2 we did not find 
a meteorite with a similar albedo value
that could match the whole VIS-NIR spectral behavior of 
92 Undina.
However, we were able to produce a model of
the surface of this M-type asteroid
with a geographical mixture of 99\% of pallasite Esquel and 1\% 
orthopyroxene.
This mixture model reproduces the spectral behavior of Undina,
however its albedo is much lower (0.14) than that of the asteroid
(0.25).

For 337 Devosa the spectral behavior has been reproduced 
by two different mixtures: one composed of 99\% pallasite
Esquel and 1\% orthopyroxene (for an albedo of 0.14) 
(red line in Fig.~\ref{modelli}), and one 
composed of 98\% pallasite Esquel and 2\% goethite (for an albedo of 0.14).
337 Devosa has spectral properties in the VIS-NIR 
spectral range that are very similar
to the M-type asteroid 22 Kalliope (see Fornasier et al. 2010). 
For these asteroids, 
the 0.9 $\mu$m band is consistent with a small
amount of anhydrous silicate (such as orthopyroxene), or with 
a small amount of goethite (an aqueous alteration product).

For 417 Suevia the proposed meteoritic analogue (the iron meteorite 
DRP78007, Fig.~\ref{fig5met}) does not fit the observed 0.9 $\mu$m band. 
We found that a geographical mixture of 97\% IM DRP78007 (RELAB file cdmb47) 
and 3\% orthopyroxene  models the asteroid spectral behaviour. 
The albedo of the mixture is 0.16, a bit lower than the value estimated for 
Suevia (0.20).
For 1124 Stroobantia the proposed meteorite analogue
(metallic meteorite DRP78007) also does not fit the faint 0.9
$\mu$m band.  For this asteroid, we propose two mixing models. One is a 
geographical mixture of 98\% IM MET101A Odessa and 2\% orthopyroxene, with 
an albedo of 0.13 (blu line in Fig.~\ref{modelli}). 
This model reproduces the asteroid spectrum below 1.4 $\mu$m 
but not the longer wavelengths.
Alternatively, a mixture of 96\% iron meteorite DRP78007 and 4\% 
olivine (albedo 0.16, red line in Fig.~\ref{modelli}) fits 
Stroobantia 
beyond 0.6 $\mu$m, but does not match the spectral region below 0.6 $\mu$m.  
Although none of the modeled mixtures reproduces all spectral features (albedo, 
spectral slopes, band depth, and band center) of the asteroid, our best inferrence from the modeling is to suggest that the surface composition of Stroobantia is probably consistent with metallic meteorites enriched in silicates.

These results show that a few percent (less than 3\%) 
of orthopyroxenes or goethite added to iron or pallasite meteorites 
can reproduce the weak spectral features around 0.9 $\mu$m seen on some asteroids 
belonging to the X-complex. 

517 Edith shows a flat and featureless spectrum. The observed spectral behavior appears 
similar to the spectrum of pure  amorphous carbon (from Zubko et al., 1996), although we 
cannot exclude the possibility that a small amount 
of silicates may be present and masked by the dark and opaque 
materials.
 
 Considering the low albedo value and the featureless and reddish spectral behavior, we classify  1328 Devota as a D-type. Since the Tagish Lake meteorite is usually 
considered the best meteorite analog for 
D-type asteroids (Hiroi et al. 2001), we
compared the spectrum of Devota with laboratory spectra of
several samples of this meteorite. 
The spectrum of Devota appears to be redder than that of Tagish Lake. 
This could be due to the different ages of the asteroids and meteorites.
The surface of Devota could be older than
the meteorite, and the reddening of the asteroid spectrum could be the result
of space weathering processes. 
Alternatively, the spectral behavior of Devota can be reproduced with 
a geographical mixture of 94\% Tagish Lake meteorite (RELAB file c1mt11)
and 6\% 
Triton tholins (from McDonald et al. 1994), resulting in an albedo of 0.02.
Our observations of Devota do not cover the wavelengths necessary for detection of the  tholin spectral feature at about 3 $\mu$m, so
although we cannot constrain the presence of 
tholins with the data in hand, 
we suggest that some tholin or tholin-like
component could be the reddening agent producing the observed 
spectral slope between 0.5 and 2.5 $\mu$m.
 
\section{The X complex: overview and discussion}

[HERE TABLE 6 and 7]

[HERE FIGURES 8 and 9]

\begin{figure*}
\includegraphics[width=14cm]{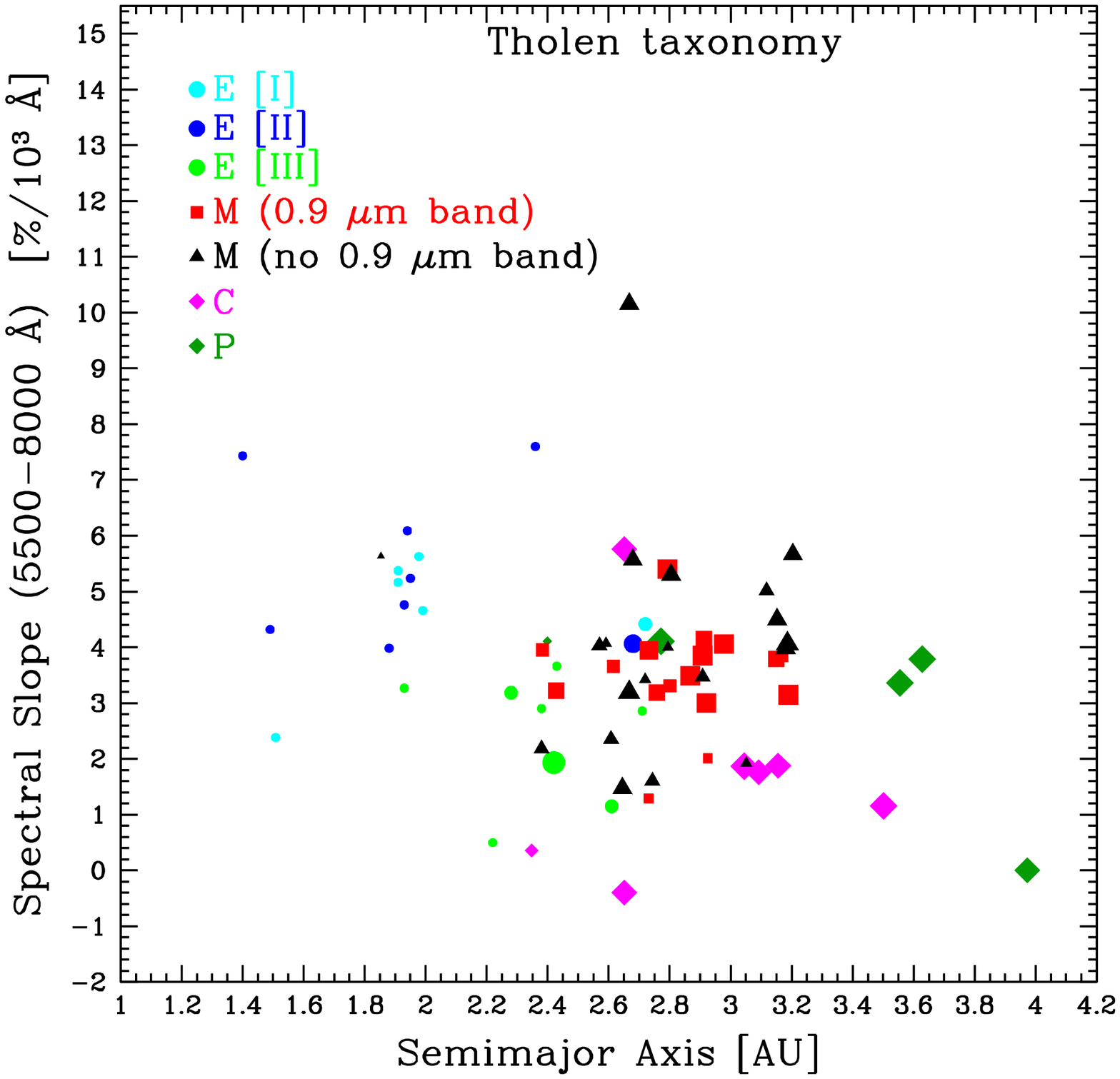}
\caption{}
\label{slopea}
\end{figure*}

Our survey devoted to the X-complex asteroids as defined by the 
Tholen (1984) taxonomy is composed of 78 objects, including the 
E and M-type asteroids already published in Fornasier et al. (2008, 2010).
On the basis of their spectral behavior and albedo values,
we classified 22 E-types and distributed them 
into 3 subclasses I, II, and III, 
(see Fornasier et al. 2008 and references therein). 
One of the objects (3447) was originally classified as X-type and 
was tentatively attributed to the E(I) class in this work, due to its moderately high albedo value but with a high uncertainty. 
Our survey includes 38 M-types, ten of which (77, 92, 184, 337, 417, 741, 758, 1124, 
1146) were originally classified as X-types, 7 C-types (498 Tokio was 
originally classified as an M-type, and all the others as X-type),
1 D (originally classified as X-type), 5 P-type (originally classified as X-type), 3 A-types (2577, 7579 and 1122 which were originally classified as E-types,
and an  X-type, respectively), and 2 S-types (5806 and 516, which were originally classified as an
E and M-type, respectively).
In Fig.~\ref{slopea} we show the spectral slope value $S_{VIS}$ calculated 
in the visible wavelength range versus the semimajor axis for all  the E, 
M and C/P-types observed. As expected, high albedo E-type asteroids populate 
mainly the Hungaria region and the inner part of the main belt, while low 
albedo C and P-types are located mainly in the outer part of the main belt 
or beyond it, and M-type are located between 2.4 and 3.2 AU. 
The mean visible spectral slope for the different E, M and P classes are 
very similar, as expected (Table~\ref{slopes}). In the NIR range, M-type asteroids having a band in the 0.9 $\mu$m region have mean spectral slope S$_{NIR1}$ and S$_{NIR2}$ values similar to P-type asteroids. These values are higher than those of the  M-type asteroids without the 0.9 $\mu$m band. For the E-type asteroids, observations in the infrared range are available only for a very few objects. We give the NIR  mean slope values only for the subtype III (4 asteroids observed in the NIR range: 44 Nysa, 214 Aschera, 317 Roxane, and 437 Rhodia), and the subtype II (64 Angelina, 2867 Steins, and 4660 Nereus), that show the lower S$_{NIR1}$ mean value within the X complex (no data are available in the NIR range for the subtype I).

One M-type asteroid, 77 Frigga, show a very peculiar spectrum, with a near infrared spectrum similar to that of other M-type bodies, but with a very high $S_{VIS}$ value comparable with that of the D-type Devosa. Nevertheless its moderate albedo value allows us to exclude any possible link with dark P or D-types. Also A or S type asteroids have $S_{VIS}$ value similar to that of Frigga, but the absence of any absorption band due to olivine and pyroxene lets us exclude any link with these asteroid classes. As discussed in section 3, Frigga may show surface variability, and the 3 $\mu$m band, usually associated with hydrated silicates, has been detected on its surface (Rivkin et al., 2000). Polarimetric data reveals that Frigga has a large inversion angle (Gil-Hutton, 2007), implying that the surface is composed of small particles (comparable to the wavelength), and/or of a mixture of particles with high contrast in albedo, like refractory inclusions seen on some carbonaceous chondrites (see Belskaya et al. 2010, and references therein).  Our attempt to find a meteorite analogue was unsuccessful as the best match found (the iron meteorite Chulafinnee) does not satisfactory reproduce the asteroid spectrum. No radar observations are available for this body. Additional observations of 77 Frigga will be very important to confirm its surface heterogeneities and possibly to constrain its surface composition. \\
In our sample 43 objects were observed in the complete 
V+NIR spectral range.
Among them 37 have spectral features and albedo values compatible with
the X-complex (5 E-types, 29 M-types, and 3 P-types).
For all objects we investigated the spectral matches  with
meteorites/minerals or geographical mixture models. 
For the 5 E-type asteroids (44, 64, 214, 317 and 437) 
we found good spectral matches with the enstatite
achondrite meteorites, in several cases enriched with troilite, oldhamite or
orthopyroxenes (Fornasier et al. 2008).
The 3 P-type asteroids presented in this paper (275, 522, and 909) 
exhibit spectral behaviors and albedo values compatible with carbonaceous chondrite meteorites. 
Most of the M-type asteroids have spectral features and albedo values well represented 
by iron meteorites, pallasites, and enstatite chondrites -- in several cases 
enriched with orthopyroxenes, olivines, or goethite.

Our new spectral observations enhance the available physical information for the observed 
asteroids and allow us to apply the  Bus-DeMeo classification recently published
(DeMeo et al., 2009). The Bus-DeMeo system is based on the
asteroids' spectral characteristics over the wavelength range 0.45 to 2.45 $\mu$m
without taking the albedo into consideration. 
In the Bus-DeMeo taxonomy, the X complex comprises 
4 types (X, Xk, Xe, Xc) as in the  Bus \& Binzel (2002) system: 
an asteroid belongs to the Xe-type if the 0.49 $\mu$m feature is present, 
to the  Xk-type if a feature is present in the 0.8--1.0 $\mu$m range, 
to the Xc-type if the spectrum is red and featureless with slight 
concave-down curvature, and to the X-type if the spectrum is straight and
featureless.
The Tholen X-type asteroids we observed show different spectral behaviors in the near infrared range. 
We therefore re-classified 16 of the asteroids presented in this paper and observed both in the VIS and NIR range according to the Bus-DeMeo taxonomy. The corresponding classes are summarized in Table
~\ref{slope}. All 5 asteroids showing the faint 0.9 $\mu$m band 
fall in the Xk-type (337 Devosa also has the 0.43 $\mu$m band), 
while 1122 Nieth, which shows a broad and deep 0.96 $\mu$m band 
and a steep infrared spectrum, is classified as an A-type.
The 5 asteroids having the 0.43 $\mu$m absorption band (note that this region is outside the wavelength limits where the Bus-DeMeo taxonomy is defined) belong 
to 5 different classes: 337 Devosa is classified as Xk; 
50 Virginia, showing bands due to hydrated materials, is a Ch; 
283 Emma is a C-type; 517 Edith is an Xc; and 1355 Magoeba, 
having also the peculiar band at 0.49 $\mu$m, is classified 
as Xe-type in the Bus-DeMeo taxonomy. 
The featureless objects belong to the X-type (77, 758, and 909), 
D-type (1328, and 1902), and to the C-type (275).

\begin{figure*}
\includegraphics[width=14cm]{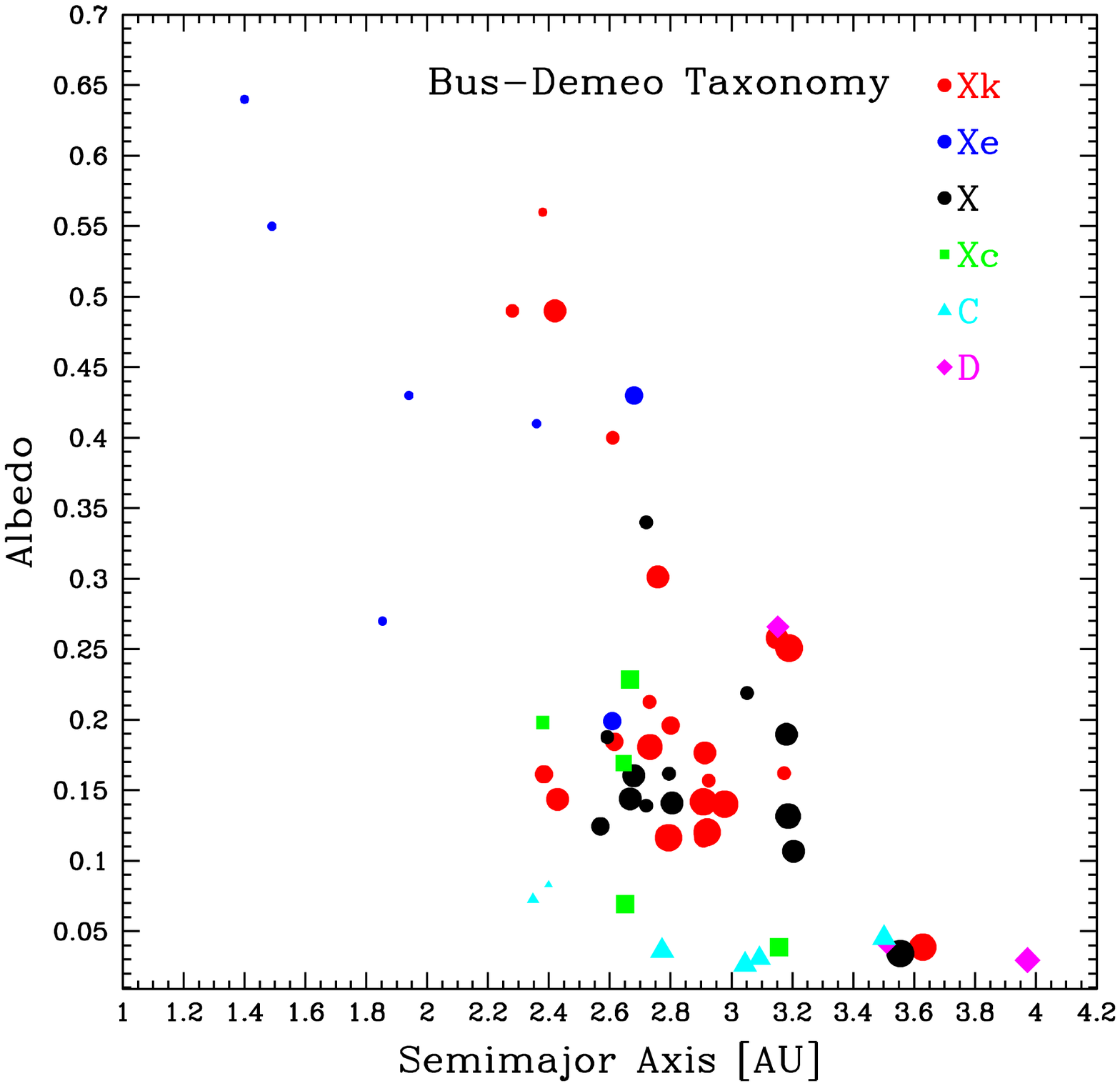}
\caption{}
\label{busalbedo}
\end{figure*}

In Table~\ref{busDeMeotx} we report Bus-DeMeo taxonomic classifications
together with the Tholen classifications (existing or proposed by us) 
for all the asteroids observed during our survey devoted to the X-complex asteroids. 
In Fig~\ref{busalbedo} we show the albedo value versus the semimajor axis 
for the observed asteroids classified with the Bus-DeMeo taxonomy. We aim 
to investigate if possible correlations between the classes of this taxonomy 
and the albedo may exist, although the albedo is not a parameter taken into 
account in the Bus-DeMeo classification system.  
We dispose of the albedo value for 7 out of the 10 Xe asteroids, that show the 0.49$\mu$m band typical of this class and attributed to sulfides. Most of them 
correspond to the high albedo subgroup II of the Tholen E-class, with only two asteroids (132 
Aertha and 1355 Magoeba) being classified as M-type in the Tholen taxonomy. So this peculiar band, attributed to the presence of sulfides like oldhamite, seems not to be exclusively associated with high albedo E-type asteroids. \\
The Bus-DeMeo Xk-class asteroids show two distinct albedo distributions: 
a high albedo group (albedo $>$ 0.4), corresponding to the subgroup III of 
the Tholen E-class, and a medium albedo group (0.1 $<$ albedo $<$ 0.3) which 
includes all the Tholen M-types showing a faint feature in the 0.9 $\mu$m 
region. The Xk-class includes also a low albedo object, 522 Helga, that we classified here as a P-type in the Tholen taxonomy. 
Iron bearing pyroxenes such as orthopyroxene are suggested to cause the feature around 0.9 $\mu$m, characteristic of the Xk class, and seem to be present on asteroids with very different albedo values. \\
Most of the Bus-DeMeo X-class asteroids have 0.1 $<$ albedo $<$ 0.2, with 
the exception of 504 Cora (albedo = 0.34, therefore classified as E[I] 
asteroid) and 909 Ulla (albedo=0.034, therefore classified as Tholen P-type).
All the C-types have low albedo values, but the Xc and D-types span low 
and moderate albedo values. In particular the asteroid 849 Ara has a steep spectral slope and falls in the D class according to the Bus-DeMeo taxonomy (Fornasier et al. 2010), but its albedo is very high (0.27), excluding a surface composition of organic-rich silicates, carbon, and anhydrous silicates as commonly expected on low-albedo Tholen D-type asteroids.

From this comparison, it is evident that the Bus-DeMeo taxonomy is very helpful in constraining the asteroids' surface compostion (for example Xk-type means presence of orthopyroxene), in particular when some absorption bands are present on the spectra. But it is clear that the asteroid albedo is also a very important parameter for constraining the surface properties and also their evolution in time (space weathering effects). We strongly encourage the development of a next-stage Bus-DeMeo taxonomy that includes the important albedo parameter for asteroid composition/classification.

\section{Summary}

We present new visible and near infrared spectra of 24 asteroids belonging to
the X-type as defined by Tholen \& Barucci (1989), that is an "E--M--P" type
asteroid for which albedo information was not available at the time of their
classification. The X complex in the Tholen taxonomy is comprised of
the E, M and P classes which
have very different mineralogies but which are spectrally similar with
featureless spectra in broadband visible wavelengths. 
Our observations reveal a large variety of spectral behaviors within the 
X class, and we identify  
weak absorption bands on 11 asteroids. We combine our spectra with the albedo 
values available since 2002 for the observed bodies to 
suggest new Tholen-like classifications.  We find:  1 A-type (1122), 1 D-type 
(1328), 1 E-type (possibly, 3447 Burckhalter), 10 M-types (77, 92, 184, 337, 417, 
741, 758, 1124, 1146, and 1355), 5 P-types (275, 463, 522, 909, 1902), and 6 C-types (50, 220, 223, 283, 517, and 536).
Four new M-type asteroids (92 Undina, 337 Devosa, 417 Suevia, and 1124 
Stroobantia) show a faint band in the 0.9 $\mu$m region, attributed to low 
calcium, low iron orthopyroxene. Indeed, several works based on spectral and 
radar observations show that not all the M-type asteroids have a pure metallic 
composition (Fornasier et al. 2010 and reference therein).\\
Three low albedo asteroids (50 Virginia, 283 Emma, and 517 Edith) show a weak 
band centered at 0.43 $\mu$m  that we interpreted as due to Fe$^{3+}$  
spin-forbidden transition in hydrated minerals (hematite, goethite). 
Also the medium albedo bodies 
337 Devosa and 1355 Magoeba have the same absorption.  In this case the band 
may be associated with chlorites and Mg-rich serpentines or pyroxene minerals 
such us pigeonite or augite. 50 Virginia shows also 
two absorptions centered at $\sim$ 0.69 and 0.87 $\mu$m which are typical of 
hydrated silicates. \\
We performed a search for meteorite and/or
mineral spectral matches between the asteroids observed in the visible and 
near infrared range (with published albedo values) and the RELAB 
database. \\
The best matches found for all the M-types of our sample are iron or pallasite meteorites as suggested in the literature, however we note that these meteorites often do not reproduce the faint absorption band features (such as the possible orthopyroxene absorption bands at 0.9 $\mu$m) detected on the asteroids.
We tried to constraint the asteroids' surface compositions using geographical mixing models for new M-types having the 0.9 $\mu$m feature. We found 
good spectral matches by enriching the iron or pallasite meteorites with small 
amounts ($<3\%$) of orthopyroxene or goethite.   
For the low albedo asteroids we found as the best match CM carbonaceous 
chondrites, either unaltered or altered 
(submitted to heating episodes or laser irradiation).
Our sample includes also two objects whose spectra diverge completely from 
other X complex asteroids: 1328 Devota, a low albedo body with 
a red featureless spectrum, suggested to belong to the D-type, and 1122 Nieth, 
a medium albedo asteroid with a broad and deep 0.9 $\mu$m band and a steep 
infrared spectrum, suggested to belong to the A-type. A 
synthetic model made with the Tagish 
Lake meteorite, considered as the best meteorite analogue of D-type asteroids 
(Hiroi et al 2001), and a reddening agent (Triton tholin) reproduces the 
spectral behavior of 1328 Devota.

The whole sample of asteroids included in our work is 72 X-type
objects (we exclude the A, D and S/Sq-type asteroids), partly published here and partly  
already published in Fornasier et al. (2008) and Fornasier et al. (2010). The analysis of this complete sample 
has clearly shown that, although
the mean visible spectral slopes of M-, E- and P-type asteroids
are very similar to each other, the differences in albedo indicate major differences in mineralogy and composition.

\bigskip

{\bf Acknowledgment} \\
The authors thank Dr. M. Ockert-Bell, Dr. M. Shepard, and 
Dr. A. Migliorini for their help with observations, and A. W. Rivkin and an anonymous referee for their 
comments and suggestions.
BEC thanks Jonathan Joseph for computer programming assistance.
SF thanks Dr. F. DeMeo for her help in the classification 
of the observed asteroids using her taxonomy. 
Taxonomic type results presented in this work were determined, in whole or in part, using a Bus-DeMeo Taxonomy Classification Web tool by Stephen M. Slivan, developed at MIT with the support of National Science Foundation Grant 0506716 and NASA Grant NAG5-12355. 
This research utilizes spectra acquired with
the NASA RELAB facility at Brown University.

\bigskip

{\bf References} \\

Adams, J. B. 1975. Interpretation of visible and near-infrared reflectance
spectra of pyroxenes other rock-forming minerals. In Infrared and Raman
Spectroscopy of Lunar and Terrestrial Minerals (C. Karr, Jr., Ed.), pp. 90--116.
Academic Press, New York.

Belskaya, I. N., Fornasier, S., Krugly, Yu. N., Shevchenko, V. G., Gaftonyuk, N. M., Barucci, M. A., Fulchignoni, M., Gil-Hutton, R, 2010. Puzzling asteroid 21 Lutetia: our knowledge prior to the Rosetta fly-by. Astron. and Astroph. 515, A29

Birlan, M., Vernazza, P., Nedelcu, D. A., 2007. Spectral properties of nine
M-type asteroids. Astron. Astroph. 475, 747--754

Britt, D.T.,Pieters, C. M., 1988. Bidirectional reflectance properties of
iron-nickel meteorites. In: Lunar and Planetary Science Conference, 503--512

Burbine, T.H., 1998. Could G-class asteroids be the parent bodies of the CM chondrites? Meteoritics \& Plan. Sci. 33, 253--258

Burbine, T. H., Cloutis, E. A., Bus, S. J., Meibom, A., Binzel, R. P., 1998. The detection of troilite (FeS) on the surfaces of E-class asteroids.  Bull. Am. Astron. Soc. 30, 711 
 
 Burbine, T. H., McCoy, T. J., Nittler, L., Benedix, G., Cloutis, E., Dickenson, T. 2002a. Spectra of extremely reduced assemblages: Implications for Mercury. Meteorit. Planet. Sci. 37, 1233--1244

 Burbine, T.H., McCoy, T.J., Meibom, A., 2002b. Meteorite parent bodies. In
Asteroids III (Bottke, W. et al. editors), pp 653--664, Univ. of Arizona Press, Tucson 

Burns, R. G., D. J. Vaughan, R. M. Abu-Eid, M. Witner, and A. Morawski 1973.
Spectral evidence for Cr$^{3+}$, Ti$^{3+}$, and Fe$^{2+}$ rather than Cr$^{2+}$,
and Fe31 in lunar ferromagnesian silicates. In Proc. 4th Lunar Sci. Conf.
983--994

  Bus, S. J., Binzel, R. P.,  2002. Phase II of the Small Main-Belt Asteroid Spectroscopic SurveyA Feature-Based Taxonomy. Icarus 158, 146--177
  
  Busarev, V. V., 1998. Spectral Features of M-Asteroids: 75 Eurydike and 201 Penelope. Icarus 131, 32--40

Carvano, J. M., Lazzaro, D., Moth\'e-Diniz, T., Angeli, C. A., Florczak, M., 2001. Spectroscopic Survey of the Hungaria and Phocaea Dynamical Groups. Icarus 149, 173--189

Clark, R.N., Swayze G.A., Gallagher, A.J., King, T.V.V., Calvin, W.M., 1993.
U.S. Geological Survey Open File Report 93-592, http://speclab.cr.usgs.gov 

Clark, B.E., Bus, S.J., Rivkin, A.S., McConnochie, T.,
Sander, J., Shah, S., Hiroi, T., Shepard, M., 2004a. E-Type asteroid
spectroscopy and compositional modeling, JGR 109, 1010--1029

Clark, B.E., Bus, S.J., Rivkin, A.S., Shepard, M., Shah, S., 2004b. Spectroscopy
of X-type asteroids. Astron. J. 128, 3070--3081

Clark, B.E., J. Ziffer, D. Nesvorny, H. Campins, A. S. Rivkin, T. Hiroi, M.A. Barucci, M. Fulchignoni, R. P. Binzel, S. Fornasier, F. DeMeo, M. E. Ockert-Bell, J. Licandro, T. Mothé-Diniz, 2010. Spectroscopy of B-Type Asteroids: Subgroups and Meteorite Analogs.  Journal of Geophysical Research, 115, E06005

Cloutis, E. A., Gaffey, M. J., Smith, D. G. W., Lambert, R. St. J., 1990. Metal Silicate Mixtures: Spectral Properties and Applications to Asteroid Taxonomy. J. Geophys. Res., 95, 281, 8323--8338

Dahlgren, M., Lagerkvist, C. I., 1995. A study of Hilda asteroids. I. CCD spectroscopy of Hilda asteroids. Astron. Astrophys. 302, 907--914

Dahlgren, M., Lagerkvist, C.I., Fitzsimmons, A., Williams, I. P., Gordon, M., 1997. A study of Hilda asteroids. II. Compositional implications from optical spectroscopy. Astron. Astrophys. 323, 606--619

DeMeo, F. E., Binzel, R P., Slivan, S. M., Bus, S. J., 2009. An extension of the Bus asteroid taxonomy into the near-infrared. Icarus 202, 160--180

Fieber-Beyer, S. K., Gaffey, M. J., Hardersen, P. S., 2006. Near-Infrared
Spectroscopic Analysis of Mainbelt M-Asteroid 755 Quintilla. 37th Annual Lunar
and Planetary Science Conference, March 13-17, 2006, League City, Texas,
abstract no.1315

Fornasier, S., Lazzarin, M., Barbieri, C., Barucci, M.A., 1999. Spectroscopic
comparison of aqueous altered asteroids with CM2 carbonaceous chondrite
meteorites. Astron. Astrophys. 135, 65--73

Fornasier, S., \& Lazzarin, M., 2001. E-Type asteroids: Spectroscopic investigation on the $0.5 \mu$m absorption band. Icarus 152, 127--133

Fornasier, S., Dotto, E., Marzari, F., Barucci, M.A., Boehnhardt, H., Hainaut, O.,
de Bergh, C., 2004. Visible spectroscopic and photometric survey of L5 Trojans
: investigation of dynamical families. Icarus 172,  221--232

Fornasier, S., Dotto, E., Hainaut, O., Marzari, F., Boehnhardt, H., de Luise, F., Barucci, M. A., 2007a.
Visible spectroscopic and photometric survey of Jupiter Trojans: Final results on dynamical families. Icarus 190, 622--642

Fornasier, S., Marzari, F., Dotto, E., Barucci, M. A., Migliorini, A., 2007b.
Are the E-type asteroids (2867) Steins, a target of the Rosetta mission,
and NEA (3103) Eger remnants of an old asteroid family? Astron. Astroph. 474, 29--32

Fornasier, S., Migliorini, A., Dotto, E., Barucci, M. A., 2008. Visible and near infrared spectroscopic investigation of E-type asteroids, including 2867 Steins, a target of the Rosetta mission. Icarus 196, 119--134

Fornasier, S., Clark, B.E., Dotto, E., Migliorini, Ockert-Bell, M., Barucci, M. A., 2010. Spectroscopic survey of  M--type asteroids. Icarus 210, 655--673

Gaffey,  M. J., 1976. Spectral reflectance characteristics of the meteorite classes. J. Geophys. Res. 81, 905--920

Gaffey, M. J., McCord, T. B., 1978. Asteroid surface materials - Mineralogical characterizations from reflectance spectra.
Space Sci. Reviews 21, 555--628

  Gaffey, M. J., J. F. Bell, and D. P. Cruikshank 1989. Reflectance spectroscopy and aster
oids surface mineralogy. In Asteroids II (R.P. Binzel, T. Gehrels, and M. S. Matthew
s, Eds.),  pp. 98--127, Univ. of Arizona Press, Tucson. 

  Gaffey, M. J., K. L. Reed, and  M. S. Kelley 1992. Relationship of E-type Apollo asteroi
ds 3103 (1982 BB) to the enstatite achondrite meteorites and Hungaria asteroids. Icarus 100, 95--109 

Gaffey, M. J., Burbine, T. H., Piatek, J. L., Reed, K. L., Chaky, D. A., Bell, J. F., Brown, R. H., 1993. Mineralogical variations within the S-type asteroids class. Icarus 106, 573--602

 Gaffey, M. J., Cloutis, E.A., Kelley, M. S., Reed, K. L., 2002. Mineralogy of Asteroids. In
Asteroids III (Bottke W. et al. editors), pp. 183-204, Univ. of Arizona Press, Tucson.

Gil-Hutton, R., Lazzaro, D., Benavidez, P., 2007. Polarimetric observations of Hungaria asteroids. Astron. Astrophys. 468, 1109--1114

Gil-Hutton, R., 2007. Polarimetry of M-type asteroids. Astron. Astrophys. 468, 1127--1132

Gil-Hutton, R., Licandro, J., 2010. Taxonomy of asteroids in the Cybele region from the analysis of the Sloan Digital Sky Survey colors. Icarus 206, 729--734

Hardersen, P. S., Gaffey, M. J., Abell, P. A., 2005. Near-IR spectral evidence
for the presence of iron-poor orthopyroxenes on the surfaces of six M-type
asteroids. Icarus 175, 141--158

Hazen, R. M., P. M. Bell, and H. K. Mao 1978. Effects of compositional variation
on absorption spectra of lunar pyroxenes. In Proc. 9th Lunar Planet. Sci. Conf.,
2919--2934

Hiroi, T., Zolensky, M. E., Pieters, C. M., 1996. Thermal metamorphism of the C, G, B, and F asteroids seen from the 0.7 micron, 3 micron and UV absorption strengths in comparison with carbonaceous chondrites. Meteoritics and Planet. Sci. 31, 321--327

Hiroi, T., Zolensky, M. E., Pieters, C.M., 2001. The Tagish Lake Meteorite: A Possible Sample from a D-Type Asteroid. Science 293, 2234--2236

Jones, T. D., Lebofsky, L. A., Lewis, J. S.,  Marley, M. S., 1990. 
The composition and origin of the C, P, and D asteroids: Water
as a tracer of thermal evolution in the outer belt. Icarus, 88, 172--192.

Khare, B. N., Sagan, C., Arakawa, E. T., Suits, F., Callcott, T. A.,
Williams, M. W., 1984. Optical constants of organic tholins produced in a 
simulated Titanian atmosphere - From soft X-ray to microwave frequencies. 
Icarus 60, 127--137.

Khare, B.N., Thompson, W.R., Sagan, C., Arakawa, E. T., Meisse, C.,
Gilmour, I. 1991. Optical Constants of Kerogen from 0.15 to 40 micron:
Comparison with Meteoritic Organics. In Origin and Evolution
of Interplanetary Dust, IAU Colloq. 126, eds. A.C. Levasseur-Regours and H.
Hasegawa, Kluwer Academic Publishers, ASSL 173, 99.

King, T.V.V., Clark, R.N., 1989. Spectral characteristics of chlorites and
Mgserpentines using high resolution reflectance spectroscopy. J. Geophys. Res. 94,
13997--14008.

Lupisko, D. F., Belskaya, I. N., 1989. On the surface composition of the M-type asteroids. Icarus 78, 395--401.

Magri, C. Nolan, M. C.; Ostro, S. J., Giorgini, J. D., 2007. A radar survey of
main-belt asteroids: Arecibo observations of 55 objects during 1999 2003. Icarus
186, 126--151.

Margot, J.-L., Brown, M.E., 2003. A low-density M-type asteroid in the mainbelt.
Science 300, 1939--1942.

McDonald, G.D., Thompson, W.R., Heinrich, M., Khare, B.N., Sagan, C. 1994.
Chemical investigation of Titan and Triton tholins. Icarus 108, 137--145.

Moroz, L. V., Hiroi, T., Shingareva, T. V., Basilevsky, A. T., Fisenko, A. V., Semjonova, L. F., Pieters, C. M., 2004.
Reflectance Spectra of CM2 Chondrite Mighei Irradiated with Pulsed Laser and Implications for Low-Albedo Asteroids and Martian Moons. Lunar Planet. Sci. Conf. 35, abstract 1279

Moth\'e-Diniz, T., Carvano, J. M., Lazzaro, D. 2003. Distribution of taxonomic classes in the main belt of asteroids. Icarus, 162, 10--21.

Ockert-Bell, M.E., Clark, B.E., Shepard, M.K., Rivkin, A.S., Binzel, R.P.,
Thomas, C.A., DeMeo, F.E., Bus, S.J., Shah, S., 2008. Observations of X/M
asteroids across multiple wavelengths. Icarus 195, 206--219.

Ockert-Bell, M.E., Clark, B.E., Shepard, M.K.,Isaacs, R.A, Cloutis, E.A., Fornasier, S., Bus, J.S., 2010.
The composition of M-type asteroids:synthesis of spectroscopic and radar observations. Icarus 210, 674--692.

Ostro, S. J., Campbell, D. B., Chandler, J. F., Hine, A. A., Hudson,
R. S., Rosema, K. D., Shapiro, I. I., 1991. Asteroid 1986
DA: Radar evidence for a metallic composition. Science, 252,
1399--1404.

Ostro, S. J., Hudson, R. S., Nolan, M. C., Margot, J. L., Scheeres,
D. J., Campbell, D. B., Magri, C., Giorgini, J. D., Yeomans,
D. K., 2000. Radar observations of asteroid 216 Kleopatra.
Science, 288, 836--839.

Pieters, C. 1983. Strength of mineral absorption features in the transmitted
component of near-infrared reflected light: First results from RELAB, J.
Geophys. Res., 88, 9534--9544.

Rivkin, A. S., Howell, E. S., Britt, D. T., Lebofsky, L. A., Nolan,
M. C., Branston, D. D., 1995. 3-micron spectrophotometric
survey of M- and E-class asteroids. Icarus, 117, 90--100.

Rivkin, A.S., Howell, E.S., Lebofsky, L.A., Clark, B.E., Britt, D.T., 2000.
The nature of M-class asteroids from 3-μm observations. Icarus 145, 351--
368.

Rivkin, A. S., Howell, E. S., Vilas, F., Lebofsky L. A., 2002.
Hydrated minerals on asteroids: The astronomical record. In
Asteroids III (Bottke W. et al. editors), pp 235--253, Univ. of Arizona Press, Tucson.

Shepard, M. K., Clark, B. E., Nolan, M. C., Howell, E. S., Magri, C., Giorgini,
J. D., Benner, L. A. M., Ostro, S. J., Harris, A. W., Warner, B., et al., 2008.
A radar survey of M- and X-class asteroids. Icarus 195, 184--205.	

Shepard, M. K., Clark, B. E., Ockert-Bell, M., Nolan, M. C., Howell, E. S., Magri, C., Giorgini,
J. D., Benner, L. A. M., Ostro, S. J., Harris, A. W., Warner, B., Stephens, R. D., Mueller, M., 2010. A radar survey of M- and X-class asteroids II. Summary and synthesis. Icarus 208, 221--237. 

Shingareva, T. V., Basilevsky, A. T., Fisenko, A. V., Semjonova, L. F., Korotaeva, N.N.., 2004. 
Mineralogy and Petrology of Laser Irradiated Carbonaceous Chondrite Mighei. Lunar Planet. Sci. Conf. 35, abstract 1137

Taylor, S. R., 1992.  Solar system evolution: a new perspective. an inquiry into the chemical composition, origin, and evolution of the solar system. In Solar System evolution: A new Perspective, Cambridge Univ.Press. 

Tedesco, E.F., P.V. Noah, M. Moah, and S.D. Price, 2002. The supplemental IRAS
minor planet survey. The Astronomical Journal 123, 10565--10585.

Tholen, D.J., 1984. Asteroid taxonomy from cluster analysis of photometry.
Ph.D. dissertation, University of Arizona, Tucson.

Tholen, D.J., Barucci, M.A., 1989. Asteroids taxonomy. In: Binzel, R.P.,
Gehrels, T., Matthews, M.S. (Eds.), Asteroids II. Univ. of Arizona Press,
Tucson, pp. 298--315. 

Vernazza, P., Brunetto, R., Binzel, R. P., Perron, C., Fulvio, D., Strazzulla,
G., Fulchignoni, M., 2009. Plausible parent bodies for enstatite chondrites and
mesosiderites: Implications for Lutetia's fly-by. Icarus 202, 477-486

Vilas, F., Hatch, E.C., Larson, S.M., Sawyer, S.R., Gaffey,
M.J., 1993. Ferric iron in primitive asteroids - A 0.43-$\mu$m absorption feature. Icarus 102, 225--231

Vilas, F., Jarvis, K.S., Gaffey, M.J., 1994. Iron alteration minerals in the visible and near-infrared spectra of low-albedo asteroids. Icarus 109, 274--283 

Warner, B. D., Harris, A. W., Vokrouhlický, D., Nesvorný, D., Bottke, W. F., 2009. Analysis of the Hungaria asteroid population.
Icarus 204, 172--182

Zellner, B., M. Leake, J. G. Williams,  Morrison, D., 1977. The E asteroids 
and the origin of the enstatite achondrites. Geochim. Cosmochim. Acta 41, 1759--1767.

Zubko, V. G., Mennella, V., Colangeli, L., Bussoletti, E., 1996. Optical 
constants of cosmic carbon analogue grains - I. Simulation of clustering 
by a modified continuous distribution of ellipsoids. MNRAS 282, 1321--1329.

\newpage

{\bf Tables}
{\scriptsize
       \begin{center}
     \begin{longtable} {|l|l|c|c|c|c|c|c|l|}  
\caption[]{Observational circumstances for the observed X type asteroids. Solar analog stars named "hip" come from the Hipparcos catalogue, "la" from the
Landolt photometric standard stars catalogue, and "HD" from the Henry Draper
catalogue}. 
        \label{tab1} \\
\hline \multicolumn{1}{|c|} {\textbf{Object    }} & \multicolumn{1}{c|}
{\textbf{Night}} & \multicolumn{1}{c|} {\textbf{UT$_{start}$}} &
\multicolumn{1}{c|} {\textbf{T$_{exp}$}} &     \multicolumn{1}{c|}
{\textbf{Tel.}} & \multicolumn{1}{c|} {\textbf{Instr.}} & \multicolumn{1}{c|}
{\textbf{Grism}} & \multicolumn{1}{c|} {\textbf{airm.}} & \multicolumn{1}{c|}
{\textbf{Solar Analog (airm.)}} \\  \hline 
\endfirsthead
\multicolumn{9}{c}%
{{\bfseries \tablename\ \thetable{} -- continued from previous page}} \\ \hline 
\endfoot
\hline \multicolumn{1}{|c|} {\textbf{Object    }} & \multicolumn{1}{c|}
{\textbf{Night}} & \multicolumn{1}{c|} {\textbf{UT$_{start}$}} &
\multicolumn{1}{c|} {\textbf{T$_{exp}$}} &     \multicolumn{1}{c|}
{\textbf{Tel.}} & \multicolumn{1}{c|} {\textbf{Instr.}} & \multicolumn{1}{c|}
{\textbf{Grism}} & \multicolumn{1}{c|} {\textbf{airm.}} & \multicolumn{1}{c|}
{\textbf{Solar Analog (airm.)}} \\  \hline 
\endhead
\hline \multicolumn{9}{r}{{Continued on next page}} \\ 
\endfoot
\hline \hline
\endlastfoot
Object  &  Night &  UT$_{start}$  & T$_{exp}$ & Tel. & Instr. & Grism & airm. & Solar Analog (airm.) \\
 & & (hh:mm) & (s) & & &   &   \\ \hline
50 Virginia & 18 Nov. 04 & 04:45 & 120 &  TNG & Dolores & LR-R & 1.09 & hyades64 (1.4) \\ 
50 Virginia & 18 Nov. 04 & 04:49 & 180 &  TNG & Dolores & MR-B & 1.10 & hyades64 (1.4) \\ 
50 Virginia & 19 Nov. 04 & 04:23 & 360 & TNG & NICS & Amici   & 1.07  & la98-978 (1.17) \\
77 Frigga  & 29 Feb. 04 & 05:12 & 90 & TNG & Dolores & LR-R    & 1.33 & la107684 (1.20)   \\
77 Frigga  & 29 Feb. 04 & 05:15 & 90 & TNG & Dolores & LR-B    & 1.34 & la107684 (1.20)   \\
77 Frigga  & 1 Mar. 04 & 05:27 & 120 & TNG & NICS & Amici   & 1.38  & la102-1081 (1.46) \\
92 Undina     &  20 Jan 07 & 04:29 & 40 & NTT & EMMI &	GR1 & 1.69 & la98-978 (1.35) \\
92 Undina     & 17 Sep. 05& 11:37 & 2560  & IRTF & SPEX &	Prism &	1.15 &
hyades64 (1.00),la115-270(1.10), \\
& & & & & & & & la93-101(1.10), la112-1333(1.10) \\
184 Dejopeja  &13 Aug 05&06:13&   300& NTT & EMMI &	GR1 &	1.07   & HD1835 (1.07) \\        	
184 Dejopeja  &12 Aug 05 & 06:59 & 360 & NTT & SOFI & GB  &1.11  & hip103579  (1.06) \\
184 Dejopeja  &12 Aug 05 & 07:04 & 360 & NTT & SOFI & GR  &1.12  & hip103579  (1.06) \\
220 Stefania & 21 Nov. 04 & 00:13 & 180 & TNG & Dolores & LR-R  & 1.12 & hyades64 (1.05)  \\
220 Stefania & 21 Nov. 04 & 00:17 & 180 & TNG & Dolores & MR-B  & 1.12 & hyades64 (1.05)  \\
223 Rosa & 20 Nov. 04 & 20:58 & 360 & TNG & Dolores & LR-R  & 1.38 & la115-271 (1.16)      \\
223 Rosa & 20 Nov. 04 & 21:05 & 420 & TNG & Dolores & MR-B  & 1.40 & la115-271 (1.16)   \\
275 Sapientia &13 Aug 05&06:44	&480& NTT & EMMI &	GR1 &	1.09  & HD1835 (1.07) \\    
275 Sapientia & 12 Jul. 04 & 08:50 & 1650 &	IRTF &	SPEX &	Prism &	  1.30 &
16CyB(1.20), la107-684(1.10)\\
 & & & & & & & &  la110-361(1.10) \\
283 Emma & 16 Nov. 04 & 04:08 & 180 & TNG & Dolores & LR-R & 1.08 & hyades64 (1.03) \\
283 Emma & 16 Nov. 04 & 04:13 &  180 & TNG & Dolores & MR-B & 1.09 & hyades64 (1.03) \\
283 Emma  & 19 Nov. 04 & 01:45 & 240 & TNG & NICS & Amici   & 1.01  & Hyades64 (1.01) \\
337 Devosa    &13 Aug 05&01:31	&300& NTT & EMMI &	GR1 &	1.08 & HD144585 (1.22) \\         	
337 Devosa    &12 Aug 05 & 03:16 & 280 & NTT & SOFI & GB &1.38  & hip083805  (1.23) \\ 
417 Suevia    &13 Aug 05 &08:12	&600& NTT & EMMI &	GR1 &	1.31  & HD1835 (1.07) \\       	
417 Suevia    &14 Aug 05 &06:04 &480& NTT & SOFI &      GB &   1.15  & la115-271 (1.15) \\
417 Suevia    &14 Aug 05 &06:20 &720& NTT & SOFI &      GR &   1.16  & la115-271 (1.15) \\	 	
463 Lola & 20 Nov. 04 & 23:48 & 180 & TNG & Dolores & LR-R  & 1.10 & hyades64 (1.05)      \\
463 Lola & 20 Nov. 04 & 23:52 & 240 & TNG & Dolores & LR-R  & 1.11 & hyades64 (1.05)      \\
517 Edith & 16 Nov. 04 & 02:42 & 180 &  TNG & Dolores & LR-R & 1.01 & hyades64 (1.03) \\
517 Edith & 16 Nov. 04 & 02:46 & 180 &  TNG & Dolores & MR-B & 1.01 & hyades64 (1.03) \\
517 Edith &  19 Nov. 04 & 03:50 & 360 & TNG & NICS & Amici   & 1.02  &  Hyades64 (1.01) \\
522 Helga     &14 Aug 05 &07:20 &600& NTT & SOFI &      GB &   1.13  & la115-271 (1.15) \\
522 Helga     &14 Aug 05 &07:38 &900& NTT & SOFI &      GR &   1.12  & la115-271 (1.15) \\
522 Helga     &  20 Jan 07 & 02:49 & 600 & NTT & EMMI &	GR1 & 1.67 & Hyades64 (1.45) \\

536 Merapi & 20 Nov. 04 & 22:23 & 150 & TNG & Dolores & LR-R  & 1.34 & la115-271 (1.28)      \\
536 Merapi & 20 Nov. 04 & 22:27 & 150 & TNG & Dolores & MR-B  & 1.34 & la115-271 (1.28)      \\
741 Botolpia  & 20 Jan 07 & 04:44 & 300 & NTT & EMMI &	GR1 & 1.70 & la98-978 (1.35) \\
758 Mancunia  &13 Aug 05 &06:05 &180 & NTT & EMMI &	GR1 &	1.03  & HD1835 (1.07) \\       	
758 Mancunia  &12 Aug 05 &06:40 &320 & NTT & SOFI &     GB &   1.05  & hip103572 (1.06) \\
758 Mancunia  &12 Aug 05 &06:46 &320 & NTT & SOFI &     GR &   1.05  & hip103572 (1.06) \\
909 Ulla       & 20 Nov. 04 & 04:41 & 180 & TNG & Dolores & LR-R  & 1.11 &  hyades64 (1.10)  \\
909 Ulla       & 20 Nov. 04 & 04:45 & 240 & TNG & Dolores & MR-B  & 1.11 &  hyades64 (1.10)  \\
909 Ulla       & 21 Nov. 04 & 05:31 & 300 & TNG & NICS & Amici   & 1.17  & la98-978 (1.17)  \\
1122 Neith & 20 Nov. 04 & 03:28 & 120 & TNG & Dolores & LR-R  & 1.14 &  hyades64 (1.10)  \\
1122 Neith  & 20 Nov. 04 & 03:31 & 120 & TNG & Dolores & MR-B  & 1.15 &  hyades64 (1.10)  \\
1122 Neith  & 21 Nov. 04 & 04:19 & 160 & TNG & NICS & Amici   & 1.32  & Hyades64 (1.23) \\
1124 Stroobantia  & 20 Nov. 04 & 03:50 & 300 & TNG & Dolores & LR-R  & 1.02 &  hyades64 (1.10)  \\
1124 Stroobantia  & 20 Nov. 04 & 03:57 & 360 & TNG & Dolores & MR-B  & 1.02 &  hyades64 (1.10)  \\
1124 Stroobantia  & 21 Nov. 04 & 04:40  & 480 & TNG & NICS & Amici   & 1.19  & la98-978 (1.19) \\ 
1146 Biarmia & 20 Nov. 04 & 20:10 & 300 & TNG & Dolores & LR-R  & 1.31 &  la115-271 (1.16)  \\
1146 Biarmia & 20 Nov. 04 & 20:19 & 300 & TNG & Dolores & MR-B  & 1.34 &  la115-271 (1.16)  \\
1328 Devota  & 20 Nov. 04 & 02:58 & 300 & TNG & Dolores & LR-R  & 1.01 &  hyades64 (1.02)  \\
1328 Devota  & 20 Nov. 04 & 03:04 & 300 & TNG & Dolores & MR-B  & 1.02 &  hyades64 (1.02)  \\
1328 Devota  & 21 Nov. 04 & 03:45 & 480 & TNG & NICS & Amici   & 1.03  & la98-978 (1.17) \\
1355 Magoeba & 16 Nov. 04 & 04:28 & 300 & TNG & Dolores & LR-R & 1.30 & hyades64 (1.03) \\
1355 Magoeba & 16 Nov. 04 & 04:34 &  360 & TNG & Dolores & MR-B & 1.31 & hyades64 (1.03) \\
1355 Magoeba &  19 Nov. 04 & 03:31 & 480 & TNG & NICS & Amici   & 1.17  & Hyades64 (1.21) \\
1902 Shaposhnikov & 14 Aug 05 & 08:03 & 480 & NTT & SOFI &    GB &   1.20  & hip113948 (1.17) \\ 
1902 Shaposhnikov &14 Aug 05 & 08:12 & 480 & NTT & SOFI &    GR &   1.25  & hip113948 (1.17) \\ 
3447  Burckhalter      & 20 Nov. 04 & 02:22 & 300 & TNG & Dolores & LR-R  & 1.42 &  hyades64 (1.02)  \\
3447  Burckhalter      & 20 Nov. 04 & 02:29 & 300 & TNG & Dolores & MR-B  & 1.46 &  hyades64 (1.02)  \\ 
\hline
\hline
\end{longtable}
\end{center}
}
\begin{list}{}{}
\item 
\end{list}

       \begin{sidewaystable}
       \caption{
Physical and orbital parameters of the  X-type asteroids
observed and selected on the basis of the Tholen taxonomy (Tholen, 1984). The Bus and Bus-DeMeo 
classifications are also reported, together with the new Tholen-like classification proposed here, 
considering the available albedo values (Tedesco et al. 2002; the albedos marked by $^{a}$ come from Gil-Hutton et al. 2007) and the spectral behaviors. The spectral slopes values
are also given (S$_{UV}$ calculated in the 0.49--0.55 $\mu$m wavelength range,
S$_{VIS}$ in the 0.55-0.8 $\mu$m range, S$_{NIR1}$ in the 1.1-1.6 $\mu$m range, S$_{NIR2}$ in the 1.7-2.4 $\mu$m
range, S$_{cont}$ in the 0.43-2.40 $\mu$m for asteroids with VIS and NIR spectra, 
and 0.43-0.92 $\mu$m for those observed only in the visible range (they are marked with an $^{*}$).  }
        \label{slope}
        \scriptsize{
\begin{tabular}{|l|c|c|c|c|c|c|c|c|c|c|c|c|c|} \hline
\hline
Asteroid & Bus &  Bus-DeMeo & NEW TX & albedo &  D (km)  & a
(AU) & e &    i$^{o}$ & S$_{UV}$ & S$_{Vis}$ & S$_{NIR1}$ & S$_{NIR2}$ & $S_{cont}$\\  \hline
50 Virginia	  & Ch &Ch  & C &0.04  &99.82   & 2.6518 & 0.2838 &2.8  & 1.57$\pm$0.74  & -0.40$\pm$0.54&  1.39$\pm$0.84& -0.16$\pm$0.77 &  1.05$\pm$0.71 \\
77 Frigga	  & X  &X   & M &0.14  &69.25   & 2.6674 & 0.1328 &2.4  & 18.23$\pm$1.32 &  10.16$\pm$0.57&  2.85$\pm$0.87&  1.08$\pm$0.83 &  5.37$\pm$0.78 \\
92 Undina	  & Xc &Xk  & M &0.25  &126.42  & 3.1895 & 0.1001 &9.9  &  6.56$\pm$0.31 & 3.15$\pm$0.54&  2.30$\pm$0.76&  1.63$\pm$0.73 &  2.46$\pm$0.71 \\
184 Dejopeja	  & X  &X   & M &0.19  &66.47   & 3.1804 & 0.0763 &1.1  & 5.05$\pm$1.02  & 3.98$\pm$0.55&  2.72$\pm$0.75&  1.38$\pm$0.75 &  2.57$\pm$0.71 \\
220 Stephania	  & C  &$-$ & C &0.07  &31.12   & 2.3482 & 0.2585 &7.5  & 2.14$\pm$0.32  & 0.36$\pm$0.55&  --&  -- &  0.74$\pm$0.72$^{*}$ \\
223 Rosa	  & C  &$-$ & C &0.03  &87.61   & 3.0920 & 0.1238 &1.9  & 3.99$\pm$0.61  & 1.76$\pm$0.57&  --&  -- &  1.86$\pm$0.73$^{*}$ \\
275 Sapientia	  &  C &C   & P &0.04   &103.00  & 2.7721 & 0.1615 &4.7  & 7.49$\pm$0.69  & 4.11$\pm$0.57&  2.21$\pm$0.75&  2.13$\pm$0.73 &  2.38$\pm$0.72 \\
283 Emma	  & $-$ &C  & C &0.03  &148.06  & 3.0447 & 0.1499 &7.9  & 2.89$\pm$0.77  & 1.87$\pm$0.54&  0.38$\pm$0.81&  0.24$\pm$0.79 &  0.73$\pm$0.72 \\
337 Devosa	  & X  &Xk  & M &0.16  &59.11   & 2.3835 & 0.1379 &7.8  & 5.57$\pm$1.28  & 3.95$\pm$0.57&  3.64$\pm$0.77&  -- &  3.58$\pm$0.02 \\
417 Suevia	  & Xk &Xk  & M &0.20  &40.69   & 2.8006 & 0.1331 &6.6  & 5.88$\pm$2.33  & 3.31$\pm$0.76&  5.94$\pm$0.76&  2.89$\pm$0.81 &  4.06$\pm$0.72 \\
463 Lola	  & $-$ &$-$& P &0.08  &19.97   & 2.3988 & 0.2202 &13.4 & 4.50$\pm$0.96  & 4.11$\pm$0.55&  --&  -- &  4.08$\pm$0.73$^{*}$ \\
517 Edith	  & C  &Xc  & C &0.04  &91.12   & 3.1558 & 0.1819 &3.1  & 2.61$\pm$0.78  & 1.88$\pm$0.54&  1.52$\pm$0.81&  0.89$\pm$0.81 &  1.59$\pm$0.71 \\
522 Helga	  & X  &Xk  & P &0.04  &101.22  & 3.6287 & 0.0755 &4.4  & 5.64$\pm$1.09  & 3.79$\pm$0.55&  3.82$\pm$0.78&  3.38$\pm$0.76 &  3.59$\pm$0.71 \\
536 Merapi	  & C  &$-$ & C &0.04  &151.42  & 3.5016 & 0.0819 &19.4 & 1.72$\pm$0.85  & 1.16$\pm$0.55&  --&  -- &  1.13$\pm$0.72$^{*}$ \\
741 Botolphia	  & X  &$-$ & M &0.14  &29.64   & 2.7198 & 0.0678 &8.4  & 6.37$\pm$0.88  & 3.42$\pm$0.54&  --&  -- &  3.22$\pm$0.74$^{*}$ \\
758 Mancunia	  & $-$ &X  & M &0.13  &85.48   & 3.1861 & 0.1518 &5.6  & 6.79$\pm$1.05  & 4.06$\pm$0.56&  2.53$\pm$0.75&  1.36$\pm$0.73 &  2.19$\pm$0.71 \\
909 Ulla	  & $-$ &X  & P  &0.03  &116.44 & 3.5550 & 0.0959&18.8 &  5.12$\pm$1.54  & 3.36$\pm$0.61&  3.20$\pm$0.89&  1.36$\pm$0.85 &  3.64$\pm$0.72 \\
1122 Nieth	  & $-$ &A  & A  & 0.45 & 12.01 & 2.6068 & 0.2562 &4.7  & 15.42$\pm$0.58  & 8.03$\pm$0.58& 15.31$\pm$1.10&  6.18$\pm$0.92 &  8.57$\pm$0.76 \\
1124 Stroobantia  & $-$ &Xk & M  &0.16  &24.65  & 2.9253 & 0.0344 &7.7 & 2.19$\pm$1.23   & 2.01$\pm$0.63&  4.73$\pm$0.88&  3.29$\pm$0.87 &  3.49$\pm$0.73 \\
1146 Biarmia	  & X  &$-$ & M &0.22  &31.14   & 3.0511 & 0.2508 &17.0 & 6.08$\pm$0.49  & 1.92$\pm$0.56&  --&  -- &  1.78$\pm$0.73$^{*}$ \\
1328 Devota	  & $-$ &D  & D &0.04  &57.11   & 3.5093 & 0.1407 &5.7   & 10.67$\pm$0.96  & 11.90$\pm$0.55& 11.70$\pm$1.10&  7.06$\pm$1.01 & 11.01$\pm$0.72 \\
1355 Magoeba	  & Xe &Xe  & M & 0.27$^{a}$  &12.90   & 1.8535 & 0.0450 &22.8  & 13.42$\pm$0.98  &  5.63$\pm$0.55&  2.23$\pm$0.81& -0.80$\pm$0.88 &  3.61$\pm$0.74 \\
1902 Shaposhnikov  & $-$ &D  & P &0.03  &96.86  & 3.9727 & 0.2242 &12.5 & --  & --&  4.08$\pm$0.77&  2.29$\pm$0.77 &  -- \\
3447 Burckhalter   & Xc &$-$ & E ? & 0.34$^{a}$  &15.60  & 1.9907 & 0.0285 &20.7 & 16.02$\pm$1.32  & 4.66$\pm$0.60&  --&  -- &  6.17$\pm$0.75$^{*}$ \\ \hline
\end{tabular}
}
\end{sidewaystable}


       \begin{table}
       \begin{center}
       \caption{
Band center, depth and width for the features detected in the
asteroid spectra.}
        \label{band}
\begin{tabular}{|l|c|c|c|c|c|c|c|c|c|c|c|c|c|c|c|} \hline
\hline
Asteroid & Band Center ($\mu$m) & Depth (\%)  & Width ($\mu$m) \\ \hline
50 Virginia	  &0.6920$\pm$0.0070  &2.2  &0.540--0.839 \\
50 Virginia	  &0.4310$\pm$0.0040  &1.1  &0.416--0.445 \\
50 Virginia	  &0.8660$\pm$0.0080  &1.2  &0.823--0.911 \\
92 Undina	  &0.9050$\pm$0.0080  &2.9  &0.742--1.027 \\
92 Undina	  &0.5103$\pm$0.0050  &0.7  &0.487--0.531 \\
283 Emma	  &0.4310$\pm$0.0040  &1.4  &0.417--0.443 \\
337 Devosa	  &0.8820$\pm$0.0080  &2.8  &0.763--1.017\\
337 Devosa	  &0.4280$\pm$0.0040  &3.0  &0.410--0.450 \\
417 Suevia	  &0.8593$\pm$0.0100  &5.2  &0.730--1.007 \\
517 Edith	  &0.4300$\pm$0.0040  &1.5  &0.416--0.444 \\
522 Helga	  &0.9400$\pm$0.0070  &2.3  &0.874--1.004\\
758 Mancunia      &1.5960$\pm$0.0060  &1.3  &1.520-1.705 \\ 
1122 Nieth	  &0.9650$\pm$0.0080  &24.3 &0.756--1.564 \\
1124 Stroobantia  &0.9170$\pm$0.0100  &5.9  &0.807--1.100\\
1355 Magoeba	  &0.4910$\pm$0.0070  &3.0  &0.443--0.543\\
1355 Magoeba	  &0.4300$\pm$0.0040  &1.9  &0.417--0.445\\ \hline
\end{tabular}
\end{center}
\end{table}

\begin{table}
       \begin{center}
       \caption{RELAB Matches (Exp means experimental).}
        \label{tab}
         \scriptsize{
\begin{tabular}{|l|c|c|c|c|l|l|} \hline
ASTEROID & ALBEDO & BEST FIT & MET    & MET  & NAME & GRAIN SIZE \\
          &       &          &  CLASS & REFL  &  & \\ \hline
50 Virginia & 0.04 & PH-D2M-032	  & CM & 0.03 & MET00639    & $<$75$\mu$m \\
77 Frigga   & 0.14 & MR-MJG-082   & IM & 0.18 & Chulafinnee &  \\
92 Undina   & 0.25 & MR-MJG-083   & IM & 0.23 & Babb's Mill & \\
92 Undina   & 0.25 & MB-TXH-043-H & Pall&0.14 & Esquel      & $<$63 $\mu$m \\
184 Dejopeja& 0.19 & MB-TXH-043-H & Pall&0.14 & Esquel      & $<$63 $\mu$m \\
275 Sapientia& 0.04 & MB-TXH-064-C   & CM Exp. & 0.03 & Murchison heated to 600$^{\circ}$C   & $<$63 $\mu$m\\
283 Emma    & 0.03 & PH-D2M-032   & CM & 0.03 & MET00639    & $<$75$\mu$m\\
337 Devosa  & 0.16 & MB-TXH-046   & IM & 0.16 & Landes      & slab \\
417 Suevia  & 0.20 & MB-TXH-047-D & IM & 0.15 & DRP78007    & 75-125$\mu$m \\
517 Edith   & 0.04 & MB-TXH-064-D & CM-Exp& 0.03 & CM Murchison  heated to 700$^{\circ}$C & 63--125$\mu$m \\ 
522 Helga   & 0.04 & MA-ATB-068   & CM Exp& 0.05 & CM Migei Laser Irrad. & $<$45$\mu$m \\ 
758 Mancunia& 0.13 & MB-TXH-046   & IM & 0.16 & Landes      & slab \\
909 Ulla    & 0.03 & MA-ATB-068   & CM Exp& 0.05 & CM Migei Laser Irrad. & $<$45$\mu$m \\ 
1124 Stroobantia& 0.16 & MB-TXH-047-A & IM & 0.12 & DRP78007 &  $<$25 \\
\hline 
\end{tabular}
}
\end{center}
\end{table}

\begin{table}
       \begin{center}
       \caption{Geographical mixing models. $^1$ blue line in Fig.~\ref{modelli},
$^2$ red line in Fig.~\ref{modelli}}
        \label{models}
         \scriptsize{
\begin{tabular}{|l|c|l|c|c|} \hline
ASTEROID & ALBEDO & GEOGRAPHICAL MODEL & FILE SOURCE & MODEL ALBEDO \\ \hline
92 Undina   & 0.25 &  99\% pallasite Esquel & RELAB ckmb43 & 0.14 \\
            &      &   1\% orthopyroxene  & RELAB cbpe40 & \\ 
337 Devosa$^1$  & 0.16 &  98\% pallasite Esquel & RELAB ckmb43 & 0.14 \\
            &      &   2\% goethite  & ASTER & \\
337 Devosa$^2$  & 0.16 &  98\% pallasite Esquel & RELAB ckmb43 & 0.14 \\
             &      &   1\% orthopyroxene  & RELAB cbpe40 & \\ 
417 Suevia  & 0.20 &   97\% iron met. DRP78007 & RELAB cdmb47 & 0.16 \\
            &      &   3\% orthopyroxene  & RELAB cbpe40 & \\ 
517 Edith   & 0.04 & 100\% amorphous carbon &  Zubko et al., 1996 & \\
1124 Stroobantia$^1$ & 0.16 & 98\% iron met. MET101A & RELAB c1sc99 & 0.13 \\
            &     & 2\% orthopyroxene & RELAB  cbpe40 & \\
1124 Stroobantia$^2$ & 0.16 & 96\% iron met. DRP78007 & RELAB cdmb47  & 0.16 \\
            &     & 4\% olivine & USGS & \\            
1328 Devota & 0.04 & 94\% carb. chond. Tagish Lake & RELAB c8mt11 & 0.03 \\
            &      & 6\% Trit. Tholin & McDonald et al. 1994  & \\ \hline
\end{tabular}
}
\end{center}
\end{table}

  \begin{table}
       \begin{center}
       \caption{Mean spectral slopes, for the different classes of asteroids with the associated standard deviation value. We discard from the following analysis a few asteroids initially classified as
E/M/X in the Tholen taxonomy but whose spectral properties diverge completely
from these classes and who have been since  reclassified as A, S or D-type
(A: 1122, 2577, and 7579; S/Sq: 516, and 5806; D: 1328).}
        \label{slopes}
\begin{tabular}{|l|l|l|l|} \hline
\hline
Asteroid Class & Mean S$_{VIS}$ & Mean S$_{NIR1}$& Mean S$_{NIR2}$ \\ 
               & (\%/($10^{3}$ \AA) & (\%/($10^{3}$ \AA) & (\%/($10^{3}$ \AA) \\ \hline
E (all)                       & 3.92$\pm$1.73  & -- & --\\
E[I]                          & 4.60$\pm$1.18  & -- & -- \\
E[II]                         & 5.15$\pm$1.31  & 0.85$\pm$0.30 &  0.32$\pm$0.15 \\
E[III]                        & 2.43$\pm$1.12  & 1.07$\pm$0.81 &  1.09$\pm$0.87    \\
M with the 0.9 $\mu$m band    & 3.54$\pm$0.90  & 3.14$\pm$1.51 &  1.82$\pm$1.42   \\
M without the 0.9 $\mu$m band & 4.00$\pm$1.97  & 1.73$\pm$2.07 &  0.81$\pm$1.21 \\
P                             & 3.34$\pm$1.69  & 2.66$\pm$1.65  & 1.83$\pm$1.25 \\
C                             & 1.77$\pm$1.94  & 0.48$\pm$0.68  & 0.17$\pm$0.35     \\ \hline
\end{tabular}
\end{center}
\end{table}

  \begin{table}
       \begin{center}
       \caption{The Tholen (existing or proposed by this work) and the Bus--DeMeo classification for all the asteroids observed withing our spectroscopical survey of the X complex. 1 = this work, 2 = Fornasier et al. 2010, 3 = Fornasier et al. 2008 and reference therein.}
        \label{busDeMeotx}
        \scriptsize{
\begin{tabular}{|l|l|l|l|l|l|l|l|} \hline
Asteroid & Tholen & Bus-DeMeo & reference & Asteroid & Tholen & Bus-DeMeo & reference\\
16   & M  & Xk & 2       &       498  & C & Xc & 2          \\
22   & M  & Xk & 2       &   	 504  & E (I) & -- & 3      \\
44   & E (III) & Xk & 3  &	 516  & S & Sq & 2          \\
50   & C & Ch & 1        & 	 517  & C & Xc & 1          \\
55   & M  & Xk & 2       &	 522  & P  & Xk & 1         \\
64   & E (II) & Xe & 3   &	 536  & C  & C & 1         \\
69   & M  & Xk & 2       &	 558  & M  & Xk & 2        \\
77   & M  & X  & 1       &	 620  & E (III) & Xk & 3   \\
92   & M  & Xk & 1       &	 741  & M  & X  & 1        \\
97   & M  & Xc & 2       &	 755  & M  & Xk & 2        \\
110  & M  & Xk & 2       &	 758  & M  & X  & 1        \\
125  & M  & -- & 2	 &	 785  & M  & X  & 2        \\
129  & M  & Xc & 2	 &	 849  & M &  D  & 2        \\
132 & E (II) & Xe & 2 	 &	 860  & M  & X  & 2        \\
135  & M  & Xk & 2	 &	 872  & M  & Xk & 2        \\
161  & M & Xc & 2	 &	 909  & P  & X  & 1          \\
184  & M  & X  & 1	 &	 1025 & E (I) & -- & 3       \\
201  & M  & X  & 2	 &	 1103 & E (III) & Xk & 3     \\
214  & E (III) & Xk & 3	 &	 1122 & A & A & 1            \\
216  & M  & Xk & 2	 &	 1124 & M  & Xk & 1          \\
220  & C  & C & 1        &	 1146 & M  & X  & 1         \\
223  & C  & C & 1	 &	 1251 & E (III) & Xk & 3    \\
224  & M  & Xc & 2	 &	 1328 & D & D & 1           \\
250  & M  & Xk & 2	 &	 1355 & M  & Xe & 1     \\
275  & P  & C  & 1	 &	 1902 & P      & D & 1      \\
283  & C  & C  & 1	 &	 2035 & E (II) & Xe & 3     \\
317  & E (III) & Xk & 3	 &	 2048 & E (II) & Xe & 3     \\
325  & M  & X  & 2	 &	 2449 & E (I) & -- & 3      \\
337  & M  & Xk & 1	 &	 2577 &  A & A &3           \\
338  & M  & Xk & 2	 &	 2867 & E (II) & Xe & 3     \\
347  & M  & Xk & 2       &	 3050 & E (III) & Xk & 3     \\
369  & M  & Xk & 2	 &	 3103 & E (II) & Xe & 3      \\
382  & M  & -- & 2	 &	 3447 & E (I) ?  & Xc & 1      \\
417  & M  & Xk & 1	 &	 4660 & E (II) & Xe & 3      \\
418  & M  & X  & 2	 &	 5806 & S      & S & 3       \\
434  & E (II) & Xe & 3	 &	 6435 & E (I) & -- & 3      \\
437  & E (III) & Xk & 3	 &	 6911 & E (II) & Xe & 3     \\
441  & M  & X  & 2	 &	 7579 & A      & A & 3      \\
463  & P  & -- & 1	 &	 144898 & E (I) & -- & 3    \\ \hline
\end{tabular}
}
\end{center}
\end{table}

\newpage

{\bf Figure captions}

\vspace{1cm}

Fig. 1 - Visible and near infrared spectra  of X-type asteroids.

Fig. 2 - Visible and near infrared spectra of X-type asteroids.

Fig. 3 - Visible and near infrared spectra of X-type asteroids.

Fig. 4 - Visible and near infrared spectra of X-type asteroids.

Fig. 5 - Best spectral matches between the observed medium albedo 
asteroids and meteorites from the RELAB database 
(see Table~\ref{tab} for details of the meteorite samples). 
All these asteroids are re-classified as M-type 
following the Tholen classification scheme. For 92 Undina 
2 meteorites are proposed: in red the IM Babb's Mill 
and in blue the IM Esquel.

Fig. 6 - Best spectral matches between the observed low-albedo asteroids and meteorites from the RELAB database (see Table~\ref{tab} for details of the meteorite samples). 

Fig. 7 - Geographical mixing model for the asteroids 92 Undina, 337 Devosa, 417 Suevia, 517 Edith, 1124 Stroobantia, and 1328 Devota (see details of each model in Table~\ref{models}). 

Fig. 8 - Spectral slope value ($S_{VIS}$) versus the semimajor axis for the different asteroids observed, classified following the Tholen taxonomy. The size of
the symbols is proportional to the asteroids' diameter.

Fig. 9 - Albedo versus semimajor axis for the different asteroids observed, classified following the Bus-DeMeo taxonomy. The size of the symbols is proportional to the asteroids' diameter.

\newpage

{\bf Figures}


\end{document}